\pdfoutput=1
\documentclass[
aip,
amsmath,amssymb,
reprint,
Conference Proceedings
]{revtex4-1}

\usepackage{graphicx}
\usepackage{dcolumn}
\usepackage{bm}
\usepackage{subcaption}
\usepackage[utf8]{inputenc}
\usepackage[T1]{fontenc}
\usepackage{mathptmx}
\usepackage{etoolbox}
\usepackage[colorinlistoftodos,prependcaption,textsize=tiny]{todonotes}

\usepackage{enumitem}
\usepackage{subfig}

\usepackage{xr}
\makeatletter
\def\@email#1#2{
\endgroup
\patchcmd{\titleblock@produce}
 {\frontmatter@RRAPformat}
 {\frontmatter@RRAPformat{\produce@RRAP{*#1\href{mailto:#2}{#2}}}\frontmatter@RRAPformat}
 {}{}
}
\makeatother
\begin{document}

\title[Numerical methods for unraveling inter-particle potentials]{Numerical methods for unraveling inter-particle potentials in colloidal suspensions: A comparative study for two-dimensional suspensions}

\author{Clare R.~Rees-Zimmerman}
\affiliation{Physical and Theoretical Chemistry Laboratory, University of Oxford, South Parks Road, Oxford OX1 3QZ, United Kingdom}
\author{Jos{\'e} Mart{\'\i}n-Roca}
\affiliation{Departamento de Estructura de la Materia, F{\'\i}sica T{\'e}rmica y Electr{\'o}nica, Universidad Complutense de Madrid, 28040 Madrid, Spain}
\altaffiliation[Also at ] {Grupo Interdisciplinar Sistemas Complejos, Madrid, Spain}
\author{David Evans}
\affiliation{Department of Chemistry, Durham University, South Road, Durham DH1 3LE, United Kingdom}
\author{Mark A.~Miller}
\affiliation{Department of Chemistry, Durham University, South Road, Durham DH1 3LE, United Kingdom}
\author{Dirk G.~A.~L.~Aarts}
\affiliation{Physical and Theoretical Chemistry Laboratory, University of Oxford, South Parks Road, Oxford OX1 3QZ, United Kingdom}
\author{Chantal Valeriani}
\affiliation{Departamento de Estructura de la Materia, F{\'\i}sica T{\'e}rmica y Electr{\'o}nica, Universidad Complutense de Madrid, 28040 Madrid, Spain}
\altaffiliation[Also at]{Grupo Interdisciplinar Sistemas Complejos, Madrid, Spain}
\email{cvaleriani@ucm.es}

\date{18 February 2025}

\begin{abstract}
We compare three model-free numerical methods for inverting structural data to obtain interaction potentials, namely iterative Boltzmann inversion (IBI), test-particle insertion (TPI), and a machine-learning (ML) approach called ActiveNet.  Three archetypal models of two-dimensional colloidal systems are used as test cases: Weeks--Chandler--Anderson short-ranged repulsion, the Lennard-Jones potential, and a repulsive shoulder interaction with two length scales.  Additionally, data on an experimental suspension of colloidal spheres are acquired by optical microscopy and used to test the inversion methods.  The methods have different merits.  IBI is the only choice when the radial distribution function is known but particle coordinates are unavailable.  TPI requires snapshots with particle positions and can extract both pair- and higher-body potentials without the need for simulation.  The ML approach can only be used when particles can be tracked in time and it returns the force rather than the potential.  However, it can unravel pair interactions from any one-body forces (such as drag or propulsion) and does not rely on equilibrium distributions for its derivation.  Our results may serve as a guide when a numerical method is needed for application to experimental data, and as a reference for further development of the methodology itself.
\end{abstract}

\maketitle

\section{Introduction\label{sec:intro}}
Understanding the behavior of colloids, and in particular their interactions, continues to enhance our fundamental knowledge of phase transitions.\cite{hansen2013} Moreover, this knowledge plays a key role in the development of advanced materials, such as photonic crystals\cite{Joannopoulos1997,Rao2020} and gels.\cite{Roldan2013,Zaccarelli2007,Lu2008} Given the unique relation between the structure of a liquid and the underlying (effective) pair potential, it is in principle possible to obtain the interactions between the colloidal particles from the structure alone through a process of inversion.\cite{henderson1974} 

Structural information can be obtained at a statistical level from scattering data\cite{Williams2001} in the form of the structure factor, which can be Fourier transformed to yield the radial distribution function (RDF).\cite{hansen2013}  Light microscopy goes further by allowing the coordinates of individual particles to be measured and tracked.\cite{Kegel2000,Weeks2000Science,Jenkins2008,Hunter2012,Ivlev2012,Elliot2001,Dinsmore2001}  Among other things, the coordinates provide another route to the RDF by direct construction of a histogram of inter-particle distances.  This approach was successfully tested on prototypical systems of attractive and repulsive interactions, such as a colloid--polymer mixture\cite{Hunter2012,Cleaver2004} and binary hard spheres.\cite{thorneywork2014} However, special care is needed when analyzing microscopy data to avoid the potential effects of imaging artifacts on the results.\cite{rao1998,baumgartl2005}

Given such structural information, interactions have been deduced for some experimental colloidal systems by assuming a particular functional form for the interactions and fitting the adjustable parameters, such as the Asakura--Oosawa depletion potential\cite{crocker1999} or a combination of electrostatic, van der Waals and interfacial interactions.\cite{wang2024}  Another approach is approximate inversion of the structure using equations for the low-density limit,\cite{royall2007} namely the potential of mean force.  To account for correlations in less dilute cases, liquid-state theories based on the Ornstein--Zernike equation in conjunction with the hypernetted chain\cite{ascough1990} or other closures,\cite{Reatto86a,Reatto88a} can also be inverted to yield an approximate pair potential directly.  Although one-step approaches are appealing for their simplicity and speed, they rarely produce an acceptable level of accuracy, and can even return potentials with spurious features.\cite{Reatto88a}  The fundamental challenge, which has long been recognized,\cite{schommers1973} is that potentials are highly sensitive to variations in the RDF.

Considering systems in thermal equilibrium, if the RDF is available it is possible to use the well-established technique of iterative Boltzmann inversion (IBI),\cite{Muller2002,Noid2013,Jadrich2017} the output of which is an effective pair potential that is fully consistent with the RDF.  Unlike the aforementioned model-fitting approaches, IBI is a model-free inverse method: no assumptions need to be made {\it a priori} about the expected form of the potential, which is beneficial when dealing with complex experimental systems.\cite{chen2010} The iterative aspect of IBI also avoids the limitations of one-step inversion methods.  A downside of IBI, and related methods based on iteration of the inverted Ornstein--Zernike equation,\cite{Reatto86a,Reatto88a,rao1998} is the need to run a new simulation at every iteration until self-consistency is achieved.

The RDF provides statistical information on structure at the pairwise level.  However, if the coordinates of the particles are directly available, then other numerical approaches become possible to extract information on the inter-particle interactions. The test-particle insertion (TPI) method,\cite{stones2019,stones2023} allows pair and higher-body distribution functions to be measured by calculating the energy change associated with the trial insertion of additional particles. TPI has been used for a long time to calculate chemical potentials.\cite{widom1963}  More recently, it has also been successfully applied to derive pair potentials from experiments of colloidal systems,\cite{stones2019} and to coarse-grain effective potentials from detailed atomistic simulations.\cite{Nicholas2024}  Like IBI, TPI makes no {\it a priori} assumptions about the form of the potential and is therefore model-free.

Graph-network algorithms have recently been used to learn the pairwise interaction and model dynamics at the particle level, such as those reported by Han \textit{et al.}\cite{Han2022} and by Ruiz-Garcia \textit{et al.}\cite{ruiz2022learning}  The latter algorithm, ActiveNet, has been developed to learn not only the pairwise interaction forces but also one-body terms, such as the active or drag forces of self-propelled colloids and the stochastic force; in this case the particle coordinates are needed.  The method does not require a functional form for the forces to be chosen and is therefore model-free, similarly to IBI and TPI.  Moreover, for active particles undergoing Brownian motion, ActiveNet is capable of learning torques and diffusion coefficients (both rotational and translational). The method has been tested on synthetic data and successfully applied to experiments of electrophoretic self-propelled Janus particles.\cite{ruiz2022learning} However, until now, ActiveNet has not been applied to suspensions of passive colloids in equilibrium states. 

IBI, TPI, and ActiveNet can all be used to unravel the interactions present in experimental suspensions of colloids.  In this article, we test the efficiency, accuracy and relative merits of these three model-free methods for extracting two-body interactions in a selection of archetypal models of Brownian colloids in two dimensions at equilibrium.  Three of the test cases are based on synthetic data, generated by simulations with known potentials.  We consider the cases of particles that interact through short-range repulsion only,\cite{weeks1971} repulsion with an attractive well,\cite{jones1924} and repulsion characterized by two length scales.\cite{gribova2009}  Finally, we test the methods on experimental data from repulsive colloidal spheres confined to two dimensions.  In all cases, both low- and high-density suspensions are considered.

\section{Simulation and experimental methods\label{sec:numerical}}

In this section, we present the methods used to produce the test data, namely the simulation techniques, the model potentials, and the preparation and characterization of the experimental colloidal suspension.

\subsection{Numerical details: Brownian dynamics simulations}

For the simulations, we have chosen a two-dimensional suspension of Brownian particles, which provides a natural comparison to our quasi-two-dimensional experimental system (see Section \ref{sec:experimental}). The equation of motion for particle $i$ can be written as 
\begin{equation}
   \gamma \, \frac{d {\mathrm{\textbf{r}}}_i}{dt} = {\mathrm{\textbf{F}}}_i + \sqrt{2 \, k_{\mathrm{B}} T \, \gamma} \; {\pmb{\eta}}_i,
\end{equation}
where ${\mathrm{\textbf{r}}}_i=(x_i,y_i)$ is the position of the $i$-th particle, $\gamma$ is the friction coefficient, $k_{\mathrm {B}}$ is the Boltzmann constant, and $T$ is the absolute temperature.  

In the Gaussian white noise vector, ${\pmb{\eta}}(t)$, each component has zero mean, $\langle \pmb{\eta} \rangle =0$, and the vector components are delta-correlated in time with unit variance over the ensemble, $\langle {\eta}^{\alpha}(t) \, {\eta}^{\beta}(t') \rangle = \delta(t-t') \delta^{\alpha\beta}$, where $\alpha,\beta$ run over the components $x,y$. 
Finally, ${\mathrm{\textbf{F}}}_i=\sum_{j\neq i} {\mathrm{\textbf{F}}}_{ij}$ is the total force acting on  particle $i$ due to the pairwise interactions with other particles $j$.  All interactions considered in this study are conservative: ${\mathrm{\textbf{F}}}_{ij} = - \mathbf{\bm{\nabla}} u_{ij}$, with $u_{ij}$ being the inter-particle pair potential.

To simulate colloidal suspensions, we perform Brownian dynamics (BD) simulations with the open-source package LAMMPS,\cite{LAMMPS} running for $t_{\mathrm{sim}}=5\cdot 10^7$ time steps and saving every 20000 steps (corresponding to 2500 saved frames). To ensure that all analysis has been performed on equilibrated results, we only analyze the last 2000 frames.
The simulations contain $N=2500$ disk-like particles with diameter $\sigma$ in a two-dimensional box with edge $L$ and periodic boundary conditions.  The value of $L$ is chosen to obtain the desired reduced number density $\rho=N\sigma^2/L^2$ for each case.

\subsection{Model potentials}

We take three widely used pairwise potentials as known targets to test the limits of the different methods. The Weeks--Chandler--Andersen (WCA) potential \cite{weeks1971} is a 12--6 Lennard-Jones potential that is truncated at $2^{1/6}\sigma$ and shifted, resulting in a purely repulsive potential of the form 
\begin{equation}
   u_{\rm WCA}(r)= \begin{cases}4 \varepsilon\left[\left(\frac{\sigma}{r}\right)^{12}-\left(\frac{\sigma}{r}\right)^6\right]+\varepsilon, & r<2^{1 / 6} \sigma \\ 0, & r \geq 2^{1 / 6} \sigma,\end{cases}
\end{equation}
where $\varepsilon$ is the unit of energy, $\sigma$ is the diameter of the particle, and $r$ is the separation of the particle centers.

The second test case is the ubiquitous Lennard-Jones (LJ) potential\cite{jones1924} truncated and shifted at $r_{\rm{c}}=2.5 \,\sigma$,
\begin{equation*}
   u_{\rm LJ, unshifted}(r) = 4 \varepsilon\left[\left(\frac{\sigma}{r}\right)^{12}-\left(\frac{\sigma}{r}\right)^6\right],
\end{equation*}
\begin{equation}
   u_{\rm LJ}(r) =  \begin{cases} u_{\rm LJ, unshifted}(r)-u_{\rm{LJ,unshifted}}(r_{\rm{c}}), & r<r_{\rm{c}} \\  0  & r \geq r_{\rm{c}}  ,\end{cases}
\end{equation}
where all symbols have the same definitions as for the WCA potential.  The LJ potential has a local minimum at $r=2^{1/6}\sigma$.

Finally, we test the so-called repulsive shoulder potential,\cite{gribova2009} which has two characteristic length scales: a hard core and a soft shell. This potential is truncated and shifted at $r_{\rm{c}}=2.8 \,\sigma$,
\begin{equation*}
   u_{\mathrm {s,unshifted}}(r)=      \varepsilon\left(\frac{\sigma}{r}\right)^n+\frac{1}{2} \varepsilon_\mathrm{s}\left\{1-\tanh \left[k_0\left(r-\sigma_\mathrm{s}\right)\right]\right\}, 
\end{equation*}
\begin{equation}
   u_{\mathrm {s}}(r)=      \begin{cases}  u_{\mathrm {s, unshifted}}(r) -    u_{\mathrm {s, unshifted}}(r_{\rm{c}}),  &  r<r_{\rm{c}} \\  0  &   r \geq r_{\rm{c}}    \end{cases},
   \label{eq:shoulderEq}
\end{equation}
where $\varepsilon$ and $\sigma$ are the energy and diameter parameters of the hard core, respectively, $\varepsilon_{\mathrm{s}}$ and $\sigma_{\mathrm{s}}$ are the height and width of the repulsive shoulder, respectively, $n$ controls the stiffness of the repulsive core, and $k_{0}$ determines the steepness of the shoulder's decay.  Following Gribova \textit{et al.},\cite{gribova2009} we set $n = 14$, $k_{0} = 10 \sigma^{-1}$, $\sigma_{\mathrm{s}} = 2.5\sigma$, and $\varepsilon_{\mathrm{s}} = \varepsilon$.

If the thermal energy is much higher than the height of the shoulder ($k_{\rm B}T\gg\varepsilon_{\rm s}$), the interactions in the shoulder potential are dominated by the hard core (the first term of the potential), whereas the effective diameter of the particles becomes $\sigma_{\rm s}$ in the opposite regime of low temperature.

For all  quantities measured in the simulations, we use $\sigma$ as the reference length scale and $\varepsilon$ as the energy scale.  The friction coefficient $\gamma$ has dimensions of mass per unit time; taking $\gamma$ as the unit of that combination defines the unit of time to be $\tau = \gamma \sigma^2\varepsilon^{-1}$.
Other parameters used in the simulations can be found in Table~\ref{tab:parameters}.
Plots of the analytic potential energy functions will be given with the numerical results in Section \ref{sec:results} (always plotted as dashed black lines).

\begin{table}[h!]
\caption{\label{tab:parameters}Parameter values for the simulated potentials and experiments: the temperature $T$, the low and high reduced-density conditions $\rho_{\rm l}$ and $\rho_{\rm h}$, and the distance $r_{\rm{c}}$ at which each input potential is truncated.  In all cases, the cutoff distance for potential extraction is $r_{\rm{a}}=3\sigma$ and the bin width used in the inversion is $\delta r=0.01\sigma$.}
\begin{ruledtabular}
\begin{tabular}{ccccccc}
&$T$& $\rho_{\rm{l}}$ &$\rho_{\rm{h}}$   & $r_{\rm{c}}/\sigma$ \\ \hline
WCA&$\varepsilon/k_{\rm{B}}$& $0.2$& $0.6$ & $1.12$ \\
LJ&$\varepsilon/k_{\rm{B}}$& $0.2$ & $0.6$ & $2.5$ \\
Shoulder&$\varepsilon/k_{\rm{B}}$& $0.1$ & $0.227$ & $2.8$\\
        &$0.1\varepsilon/k_{\rm{B}}$& $0.1$ & $0.227$ & $2.8$\\
        &$10\varepsilon/k_{\rm{B}}$& $0.1$ & $0.227$ & $2.8$\\
Experiment& $295\,{\rm K}$ & $0.2$ & $0.6$ &N/A\\
\end{tabular}
\end{ruledtabular}
\end{table}

\subsection{Experimental system\label{sec:experimental}}

Snapshots of particle coordinates are obtained from an experimental system of $\sigma=$ 2.84 $\mu \mathrm{m}$ diameter melamine formaldehyde (MF) spherical particles (microParticles GmbH, MF-COOH-AR1128) dispersed in water. This system has previously been identified to be a good model for hard spheres.\cite{thorneywork2016}

Samples are prepared as follows: (1) A glass sample cell (Hellma Flow-Through Cuvette 137-2-40, 2 mm path length) is cleaned with a 1\% solution of Hellmanex\textregistered\ III, followed by isopropanol and deionized water (Milli-Q Direct-Q\textregistered\ 3 UV Water Purification System), and then plasma cleaned for 2 min. (2) The sample cell is then filled with the particle dispersion, and the inlet/outlet ports are sealed with Parafilm\textregistered. (3) Particles sediment to the bottom of the sample cell, forming a quasi-two-dimensional monolayer with reduced particle number density $\rho  = N\sigma^2/A$, where $A$ is the area of the sample. 

This experimental system performs very well as a quasi-two-dimensional experiment: the particles are confined by gravity into a very small gravitational height, $h_{\rm{g}}$:
\begin{equation}
h_{\rm{g}}=\frac{6 k_{\rm{B}}T}{\pi \sigma^3 g \Delta \tilde{\rho}_{\rm{p-s}}}=0.068\mu \rm{m},
\end{equation}
where $g$ is the acceleration due to gravity, and the mass density difference between a particle and the solvent is estimated to be $\Delta \tilde{\rho}_{\rm{p-s}}=(1510-1000)\;{\rm kg\;m^{-3}}$. The gravitational height shows how far a particle may move vertically against gravity due to its thermal energy. As it is only 2.4\% of $\sigma$, the particles are sufficiently confined to closely resemble a two-dimensional system.\cite{mackay2024} This is important because any significant out-of-plane displacements would alter $g(r)$ relative to the strict two-dimensional case, leading to artifacts in the inferred $u(r)$, such as an apparent attraction.\cite{rao1998}

Once the sample is ready, images are taken at one frame per second with a Nikon Eclipse Ti2 inverted microscope (CFI S Plan Fluor ELWD 40XC lens \& XIMEA camera xiQ USB 3.0 SuperSpeed, $2048\times2048$ pixels) for a total of 2000 images for each density. 
Particle coordinates are located using Trackpy in Python.\cite{trackpy} The coordinate frames are cropped to the central $1648\times1648$ pixels, where the image quality is the best. 

An average of 2250 and 6756 particles were present in each low ($\rho_{\rm{l}}$) and high ($\rho_{\rm{h}}$) density image, respectively. Note that in contrast to the simulations, non-periodic boundary conditions are required to analyze the experimental data.

\section{Potential extraction methods}

\begin{figure*}[ht!]
   \centering
\includegraphics[width=0.95\linewidth]{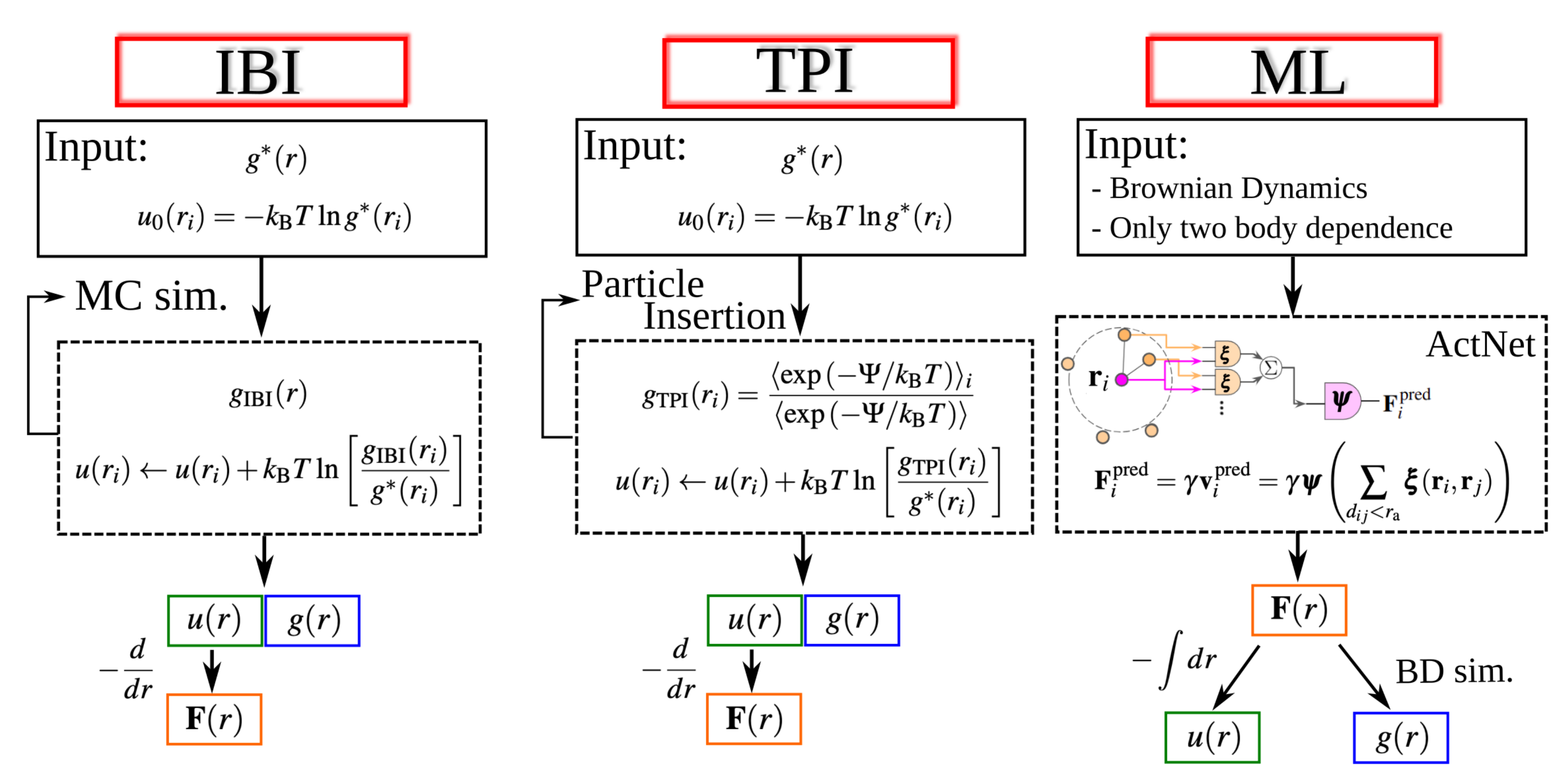}
   \caption{Schematic workflow for each of the inversion methods. IBI and TPI use a predictor-corrector scheme, with the potential of mean force as the initial guess, obtaining an updated $u(r_i)$. IBI's update scheme calculates $g_{\rm{IBI}}(r_i)$ following a MC simulation, while TPI's update scheme calculates $g_{\rm{TPI}}(r_i)$ using test-particle insertion. The ML method guesses some dynamics for the system, in this case BD, which assumes that the velocities of the particles are proportional to the forces acting on them. The GNN takes the trajectory of the $i$-th particle along with its velocities to establish the correlation between the force it feels and its velocity, as shown in Eq.~(\ref{eqn_ML}). The first network determines the relationship between the $i$-th particle and all its $j$-th possible neighbors using the edge function. Subsequently, all neighbor contributions are summed and the second network determines the relationship between the sum of all pairwise relationships and the $i$-th particle, resulting in the force on the $i$-th particle for certain coordinates relative to its neighbors. Once the network has been trained with the trajectories of all the particles, it can be used to predict the force on a particle for an arbitrary relative position. In this way, ML directly obtains $\rm{\textbf{F}}(\it{r})$, which requires integration to get $u(r)$.  A new simulation (MC using $u(r)$ or BD using $\mathbf{F}(r)$) or TPI sampling (using $u(r)$) is needed to obtain $g(r)$ for the predicted interactions.}
   \label{fig:methods}
\end{figure*}

Here we describe the three methods for extracting information on interactions from the simulated and experimentally measured data: IBI, TPI and ActiveNet (referred to as ML for machine learning).  The methods are summarized schematically in Fig.~\ref{fig:methods}.

\subsection{Iterative Boltzmann inversion}

Iterative Boltzmann inversion is typically used in bottom-up coarse-graining to derive an effective potential starting from the radial distribution function $g^{*}(r)$ derived from the target system.\cite{mcgreevy1988,Soper96a,reith2003} The inversion of the RDF to obtain a potential relies on Henderson's theorem,\cite{henderson1974} which shows that the potential underlying a given RDF at a specified temperature and density is unique, up to an additive constant, provided that only pairwise interactions are at play.  This means that if IBI can find a potential that produces the same radial distribution function as the target system, then it must be the ``true'' potential.

Coarse-graining is not the only possible application of IBI, as it can also be used to analyze experimental systems if the RDF can be obtained.  Indeed, the same underlying idea was used to extract effective pair potentials for liquid metals from experimentally derived RDFs long before the IBI name was coined.\cite{schommers1973,Schommers1983}  The RDF can be obtained by Fourier transform of the structure factor, as in those early applications, or directly by measuring inter-particle distances from confocal microscopy images in the colloidal case. IBI has also recently been used to estimate effective pairwise interactions between active (self-propelled) Brownian particles at low activity, despite the fact that active particles are out of equilibrium.\cite{evans2024}. 
 
IBI permits any shape of potential to emerge, up to a specified resolution $\delta r$, by representing $u(r)$ in piecewise linear form.  This amounts to discretizing $u(r)$ on a grid of points $r_i=(i+\frac{1}{2})\delta r$ with index $i=0,1,2\dots$.  Intermediate values of $u(r)$ are then obtained by linear interpolation.  We always choose the resolution of $u(r)$ to match the bin width of the RDF, which we take to be $\delta r=0.01\sigma$ for all test models.  Hence, $i$ is also the index of an RDF histogram bin and $r_i$ lies at the center of bin $i$.

IBI compares the target RDF $g^{*}(r)$ of the system under study to the RDF $g_{\rm IBI}(r)$ produced by a Monte Carlo (MC) simulation using a trial potential $u(r)$.\cite{allen2017}  The deviations of $g_{\rm IBI}(r)$ from $g^{*}(r)$ are then used to provide a new, improved trial potential by adjusting each discretized point according to Schommers' scheme,\cite{schommers1973}
\begin{equation}
   u(r_i) \leftarrow u(r_i) + k_{\textrm{B}}T \ln \left[\frac{g_{\rm IBI}(r_i)}{g^{*}(r_i)}\right].
   \label{eq:IBIupdate}
\end{equation}
We must choose a cutoff $r=r_{\rm a}$ beyond which we assume $u(r)=0$.  For all methods in this work, we take $r_{\rm{a}}=3\sigma$.  Note that the cutoff $r_{\rm{a}}$ for the analysis does not need to be the same as the cutoff $r_{\rm{c}}$ used for the potentials in the original BD simulations. Indeed, if just given RDF data to analyze, we would not know  {\it a priori} what $r_{\rm{c}}$ was.

Performing a MC simulation with the updated trial potential from Eq.~(\ref{eq:IBIupdate}) should result in the RDF $g_{\rm IBI}(r)$ coming closer to the target, $g^{*}(r)$.  Further iterations of Eq.~(\ref{eq:IBIupdate}), alternating with revised MC simulations, are repeated until satisfactory agreement between $g_{\rm IBI}(r)$ and $g^{*}(r)$ is obtained.

An initial guess of $u(r)$ is required before the first iteration of IBI.  In this work, we use the potential of mean force
\begin{equation}\label{initguess}
   u(r_i) = - k_{\textrm{B}}T \ln g^{*}(r_i)
\end{equation}
as the starting point.

For all results in this paper, MC simulations (in the $NVT$ ensemble) are performed with 500 particles, equilibrated for 50,000 sweeps and a main run of $10^{8}$ sweeps, saving configurations every $10^{3}$ sweeps. Hence, $10^{5}$ independent frames are used to calculate the RDF.

\subsection{Test-particle insertion}\label{subsecTPI}

The TPI inversion method exploits the fact that the RDF can be found from sample configurations of a system not only by constructing the histogram of inter-particle distances but also by sampling the energy associated with trial insertions of a test particle.  TPI seeks the potential $u(r)$ that makes these two RDFs consistent.\cite{stones2019}

As in IBI, we use the potential of mean force of Eq.~(\ref{initguess}) as the initial guess of $u(r)$ and the Schommers predictor-corrector scheme\cite{schommers1973,stones2019}
\begin{equation}
u(r_{i}) \leftarrow u(r_{i})+k_{\rm B}T \ln \left[ \frac{g_{\rm{TPI}}(r_{i})}{g^{*}(r_{i})} \right],
\end{equation}
analogous to Eq.~(\ref{eq:IBIupdate}), to iterate until a converged $u_{\rm TPI}(r_i)$ is obtained, up to a cutoff distance $r_{\rm{a}}=3\sigma$.

Originating from Widom's potential distribution theorem,\cite{widom1963} $g_{\rm{TPI}}(r_i)$ is given by the ratio of the local and bulk ensemble averages,
\begin{equation}\label{eqnTPI}
g_{\rm{TPI}}(r_i) = \frac{\langle \exp \left(-\Psi/k_{\rm B}T \right) \rangle_{i}}{\langle \exp \left(-\Psi/k_{\rm B}T \right) \rangle}.
\end{equation}
Here, the additional potential energy due to the hypothetical insertion of a test particle is denoted by $\Psi$.  The local ensemble average in the numerator, denoted $\langle...\rangle_{i}$, only includes test particles in the bin at distance $r_i$ from any particle, and the bulk ensemble average in the denominator, $\langle...\rangle$, includes all test particles.

There are different computational schemes for placing the test particles when calculating the ensemble averages. Using the same test particle coordinates for successive iterations requires the least computational effort; the distances between the particles and the test particles then only need to be calculated once. In this case, $u(r_i)$ will keep updating until $g_{\rm{TPI}}(r_i) = g^{*}(r_{i})$ to machine precision, as calculated using these test particle coordinates.  Hence, the resulting $u(r)$ will faithfully reproduce any statistical noise in $g^{*}(r_i)$.  This is a result of over-fitting to the specific choice of test particle coordinates. 

An alternative method is to use random insertion points, changing with each iteration (see the supplementary material, Sec.~S1). This is  computationally more intensive, requiring recalculation of the distances every iteration. Nevertheless, it does not take as many iterations to converge to its final precision as $g_{\rm{TPI}}(r_i)$ retains some independent noise relative to $g^{*}(r_{i})$.

In all examples presented in this work, the number of trial insertions is four times the number of particles in the system.  Since the area of each BD simulation is scaled to maintain 2500 particles in the box, $10^4$ test particles are used in the TPI inversions.  For the experimental data, the number of test particles is adapted according to the mean number of particles per frame.  In each test system, the trial insertion positions are set to a uniformly spaced square grid and the locations are held fixed for all iterations. The impact of this choice is discussed further in the supplementary material, Sec.~S1.  

\subsection{ActiveNet}

Given the detailed trajectories of the particles in a system, ActiveNet\cite{ruiz2022learning} applies machine learning methods to obtain the forces acting on the individual particles: both one-body forces, ${\mathrm{\textbf{F}}_i^{(1\rm B)}}$ (arising, for example, from external forces, walls or propulsion forces), and two-body forces, ${\mathrm{\textbf{F}}_{ij}^{(2\rm B)}}$ (the pairwise interaction forces) are learned.
The method assumes that particles follow an over-damped Langevin equation that relates velocities to forces via
\begin{equation}
   \mathrm{\textbf{v}}_i^{\, \mathrm{pred}} = \frac{1}{\gamma} \, {\mathrm{\textbf{F}}_i^{(1\rm B)}} + \frac{1}{\gamma} \, \sum_{j} {\mathrm{\textbf{F}}_{ij}^{(2\rm B)}},
\end{equation}
where $\gamma$ is the friction coefficient and $\mathrm{\textbf{v}}_i^{\, \mathrm{pred}}$ is the velocity predicted by ML.  Another way of interpreting this expression is as an expansion up to second order for the velocity as a function of force.

In our examples, the one-body force is always zero, ${\mathrm{\textbf{F}}_i^{(1\rm B)}}=0$, since there are no external potentials acting on the particles and, differently from the original implementation of ActiveNet, particles are passive (rather than active or self-propelled).  Thus, in our modified version of ActiveNet, the one-body force is set to zero from the beginning and the network does not need to learn it.\footnote{ActiveNet can be used without modifications, obtaining zero for the one-body forces in the cases  we have studied in this work.  However, the full method would need greater resources to converge and would not obtain relevant additional information for this study.}  
As reported by Ruiz-Garcia \textit{et al.} \cite{ruiz2022learning} the Graph Neural Network (GNN) in ActiveNet consists of two coupled neural networks, one to learn the one-body force and another to learn the two-body forces.  The absence of one-body forces (besides the stochastic force) in our examples means that  forces that the method has to learn do not depend on the absolute positions or orientations of individual particles (as in the case of self-propelled particles).  Therefore, the fitting function that contains all  information about the network (the so-called node function), $\pmb{\psi}$, will only depend on the relative positions of pairs of particles.

\begin{figure*}[t]
\begin{center}
   \includegraphics[width=1.0\linewidth]{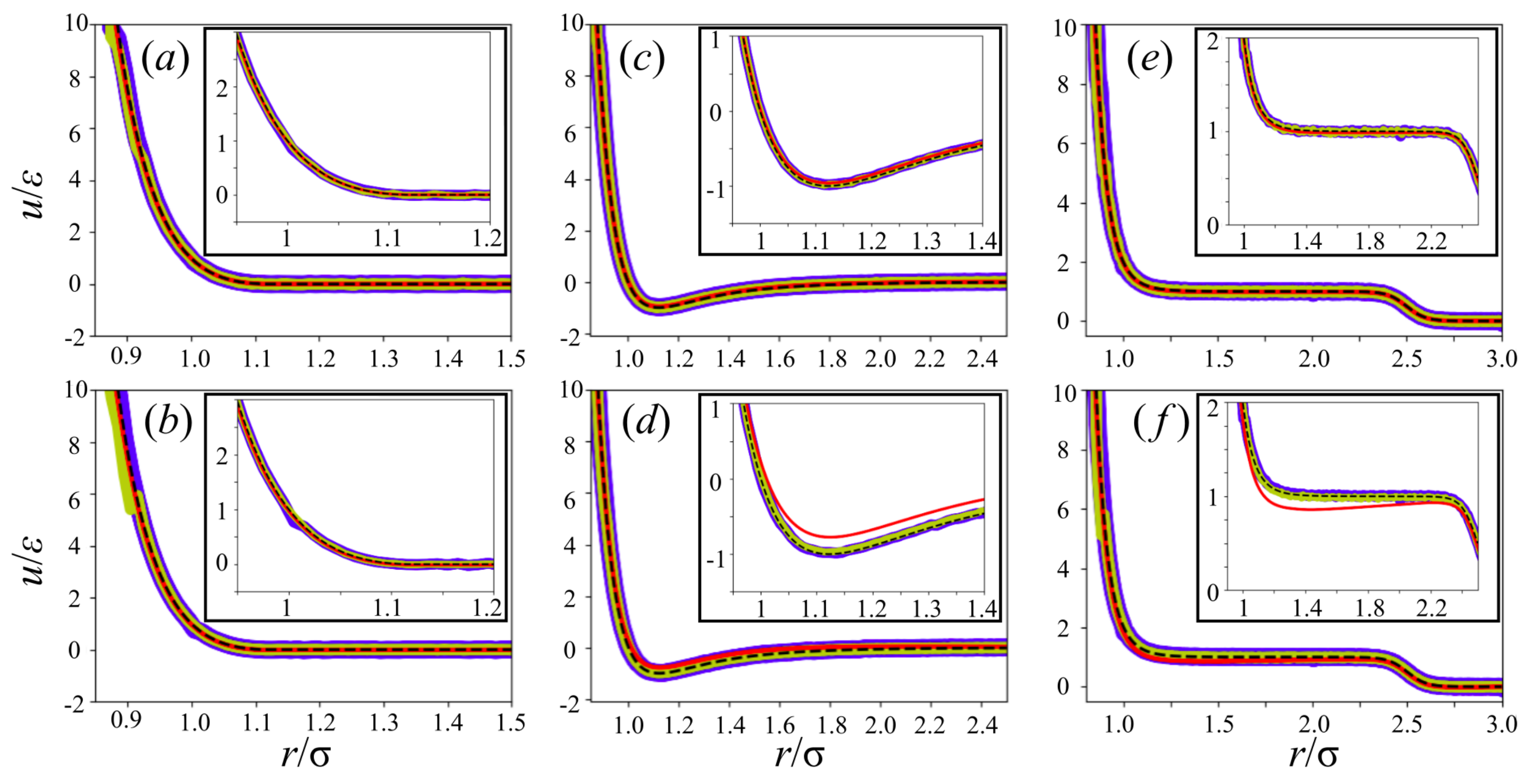}
\end{center}
   \caption{Interaction potentials $u(r)$ as a function of $r$ estimated for particles interacting via WCA at (a) low density and (b) high density, LJ at (c) low density and (d) high density, and shoulder potential at (e) low density and (f) high density. The lines have different thicknesses to aid visibility due to the large overlap between them.  From thinnest to thickest, the lines correspond to the analytic expressions (dashed black), ML (red), IBI (olive-green), and TPI (blue).}
   \label{fig:comp-simu}
\end{figure*}

Using this structure, after training, the predicted two-body force on the $i$-th particle due to the interaction with the other particles (neighbors within a cutoff distance) is given by 
\begin{equation}\label{eqn_ML}
 \mathrm{\textbf{F}}_i^{\, \mathrm{pred}} =  \gamma \,\mathrm{\textbf{v}}_i^{\, \mathrm{pred}} = \gamma \, \pmb{\psi}\left( \sum_{d_{ij}<{r_{\rm{a}}}} \pmb{\xi}(\mathrm{\textbf{r}}_i,\mathrm{\textbf{r}}_j) \right),
\end{equation}
where the node function $\pmb{\psi}$ considers the interactions between a given particle and all the others;   $\pmb{\xi}(\mathrm{\textbf{r}}_i,\mathrm{\textbf{r}}_j)$ represents the edge function, which takes as inputs the features of particles with Euclidean separation $d_{ij}=|{\bf r}_i-{\bf r}_j|$ smaller than the analysis cutoff distance, $r_{\rm{a}}=3\sigma$.  Both components of the inter-particle vectors are used as training features.  Hence, ActiveNet does not assume that the forces are isotropic.  We train the GNN with 2000 frames for the system at equilibrium and for 200 epochs. We use the same configuration of layers and neurons as set out in the original implementation by Ruiz-Garcia \textit{et al.}\cite{ruiz2022learning}

Note that ActiveNet predicts interaction forces, in contrast to IBI and TPI, which predict pair potentials. This makes it difficult to fairly compare the methods directly. However, to allow some comparison, to a first approximation we numerically integrate the force predicted by the network, obtaining the interaction potential. The force acting on a given particle predicted by the network, $\mathrm{\textbf{F}}^{\, \mathrm{pred}}(r,\theta)$, is not necessarily a central force, as assumed by the other two methods. However, we can average the angular dependence of the force over all possible orientations and project in the radial direction, $\mathrm{\textbf{F}}(r) \approx \langle \mathrm{\textbf{F}}^{\, \mathrm{pred}} \cdot \hat{\mathrm{\textbf{r}}} \rangle \; \hat{\mathrm{\textbf{r}}}$, where $r$ is the scalar distance between two particles and $\hat{\mathrm{\textbf{r}}}$ is the unit vector along the center-to-center direction. The resulting potential is given by
\begin{equation}\label{eq:ftou}
   u(r) \approx - \int_{r_{\rm{a}}}^r \mathrm{\textbf{F}}(r^{'}) \cdot \rm{d} \mathrm{\textbf{r}^{'}}.
\end{equation}
For simplicity, we will refer to the modified ActiveNet method as ML when comparing it with IBI and TPI.

\section{Results \label{sec:results}}

We now test the IBI, TPI, and ML methods for deducing the interaction potentials from data on the simulated suspensions of purely repulsive particles (WCA), short-range attractive particles (LJ) and particles with two repulsive length scales (shoulder potential), as well as on the experimental suspension of model hard sphere colloids. For the tests at $k_{\rm B}T=\varepsilon$ on the simulation data, all results are plotted in Fig.~\ref{fig:comp-simu}.

\subsection{Weeks--Chandler--Anderson particles}\label{sec4a}

WCA is a smoothly varying potential, posing little difficulty for TPI. Thus, as shown in Fig.~\ref{fig:comp-simu}(a) and (b),  TPI (blue line) easily converged to $u_{\mathrm{WCA}}(r)$ (dashed black line) for both tested densities. 
We are not concerned with the precise value of $u(r)$ returned by the inversion for $r\lesssim 0.9\sigma$, since any value of $u(r) \gg k_{\rm{B}} T$ effectively results in (approximately) no particles being predicted to reside at such distances in a simulation.  As the density increased, more iterations were required for convergence of TPI (from $\sim 40$ at $\rho_{\rm{l}} = 0.2$ to $\sim 300$ at $\rho_{\rm{h}}= 0.6$) due to the greater difference between the potential of mean force and $u_{\mathrm{WCA}}(r)$.

Similarly to TPI, IBI (olive-green line) has no difficulty in finding $u_{\mathrm{WCA}}(r)$ (dashed black line) at both low [panel (a)]  and high [panel (b)] density.  For the low-density case, convergence is reached within ten iterations.
A kink in the potential is observed around $r = 0.9\sigma$, which results from the naturally poor sampling during the MC simulation at small distances because of the relatively high potential.  This means that there are bins with few samples and high statistical uncertainty which are being compared to $g^*(r)$, which is very small, but non-zero, in this region.  Since this kink exists in a region that is inaccessible to the vast majority of particles in the simulation, we are not concerned by its presence.

Finally, the ML result (red line) nicely matches the analytical potential (black dashed line), independently of the density of the system and without fluctuations, despite the additional step of integration [Eq.~(\ref{eq:ftou})] required to yield a potential from the forces.

\subsection{Lennard-Jones particles}

The corresponding graphs for the LJ potential are presented in Fig.~\ref{fig:comp-simu}(c) and (d). Again, excellent agreement with $u_{\mathrm{LJ}}(r)$ is attained using IBI and TPI. For both densities, TPI (blue lines) easily converged to the expected $u_{\mathrm{LJ}}(r)$ (dashed black lines). The LJ potential is smooth, and at the chosen temperature of $T=\varepsilon/k_{\rm B}$, there are no regions where the density is too high (which could result in inefficient sampling due to the high energy of test insertions). Again, the high-density case [panel (d)] took more iterations to converge due to the greater difference between the initial guess of the potential of mean force and the true $u_{\mathrm{LJ}}(r)$. Convergence is obtained with marginally fewer iterations than were required for the WCA case.

IBI (olive-green lines) also has no issues in converging to the correct LJ potential for both high [panel (d)] and low [panel (c)] densities.  Similar to the WCA case, at low density convergence is reached within ten iterations, but more iterations (70) are required for convergence in the high-density case.

Finally, for low density [panel (c)], ML (red lines) produces good agreement with the analytical expression for the potential (black dashed line).  However, we observe a significant deviation in the results around the minimum of the potential when the density is high [panel (d)]. At least two factors conspire to produce this discrepancy. First, the ML method predicts the interaction force; so when integrating it numerically in order to obtain the potential, small discrepancies between the analytical force and the predicted force accumulate and are prominently reflected in the potential.  Second, the reason for the deviations in the underlying force is that the autocorrelation time of the velocity is drastically reduced at high density.  This is a critical factor for the measurement of the interaction potential using this particular tool, as discussed in the supplementary material, Sec.~S3.

\subsection{Shoulder potential particles}\label{sec4c}

Figures \ref{fig:comp-simu}(e) and (f) present the results for the shoulder potential at $k_{\rm B}T=\varepsilon$. IBI (olive-green line) and TPI (blue line) once more attain very good agreement with $u_{\mathrm{s}}(r)$ (black dashed line). As in the case of the WCA, the potentials produced by IBI have a small artifact at $r \approx 0.9 \sigma$ due to sparse sampling in the steeply repulsive core, but this is not a matter of concern in practice.  The ML method (red line) does not fully capture the shoulder potential at high densities [panel (f)], but it approximates the analytical potential at low densities [panel (e)] very well. Again, this discrepancy is due to the loss of velocity correlations when the density is high, which could be mitigated in principle, by using a smaller time interval between snapshots in the simulations for these cases (see discussion in the supplementary material, Sec.~S3). However, we should stress that the ML method was developed to learn interaction forces, not potentials.\cite{ruiz2022learning}

\begin{figure}[h!]
   \centering
   \includegraphics[width=0.95\linewidth]{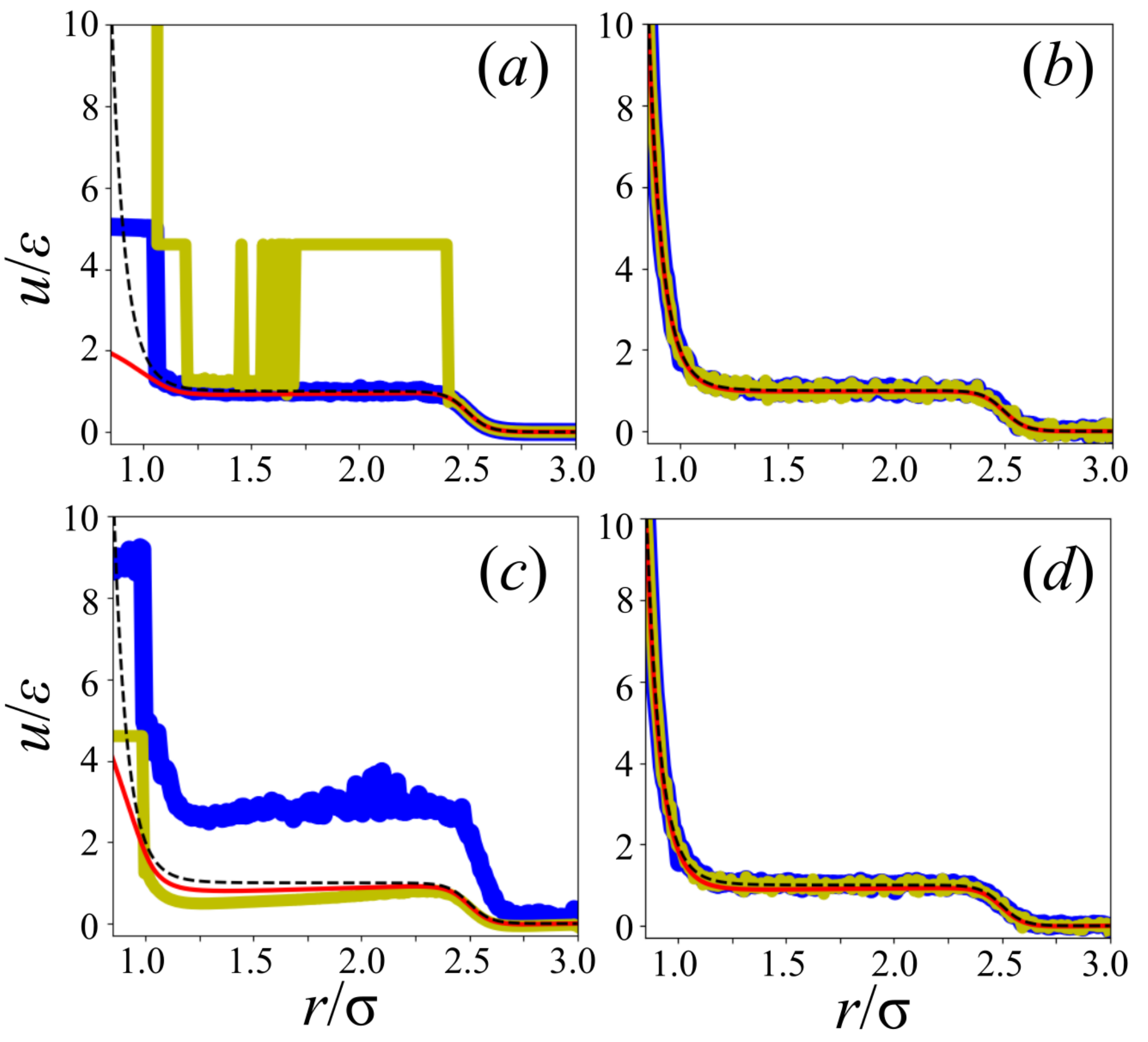}
   \caption{Numerical shoulder potential from IBI (olive-green), TPI (blue), and ML (red) for $\varepsilon_{\mathrm{s}}=\varepsilon$ and different temperatures and densities: panel (a) $\rho=0.1$, $k_{\rm{B}}T/\varepsilon=0.1$; panel (b) $\rho=0.1$, $k_{\rm{B}}T/\varepsilon=10$; panel (c) $\rho=0.227$, $k_{\rm{B}}T/\varepsilon=0.1$; and panel (d) $\rho=0.227$, $k_{\rm{B}}T/\varepsilon=10$. The black dashed line represents the analytical potential from Eq.~(\ref{eq:shoulderEq}). }
   \label{fig:shoulder}
\end{figure}

Different temperatures pose a greater challenge for inversion.  If $k_{\rm B}T\ll\varepsilon_{\rm s}$, then particles will rarely surmount the barrier onto the shoulder through thermal fluctuations, but may be forced onto it if the density is high enough.  In the opposite regime, $k_{\rm B}T\gg\varepsilon_{\rm s}$, the structure is dominated by the main repulsive core, and there is a risk that structural signatures of the shoulder are washed out.  Tests for $k_{\rm B}T/\varepsilon=0.1$ and $10$ are presented in Fig.~\ref{fig:shoulder}.

In the low-temperature case, $k_{\rm{B}}T/\varepsilon = 0.1$, at low density [Fig.~\ref{fig:shoulder}(a)], the lack of particles in the shoulder region means that $g(r)\approx0$, and therefore there is little information about the range $r < 2.5\sigma$.  Nevertheless, TPI (blue line) manages to correctly obtain $u(r)$ beyond the hard core.  The shape of the core repulsion itself is not well captured, but the steep rise in $u(r)$ at short range is qualitatively represented.  In the high-density case, at $k_{\rm{B}}T/\varepsilon = 0.1$ [Fig.~\ref{fig:shoulder}(c)], a larger proportion of particles are forced onto the shoulder due to the reduced area per particle.  However, there is still a lack of information in $g(r)$ for $2.0 < r/\sigma < 2.6$.  Although TPI (blue line) does not now quantitatively reproduce $u_{\mathrm s}(r)$, the two repulsion length scales are still clear. The TPI inversion returns $u(r) \approx 3\varepsilon$ rather than $\varepsilon$ for $r < 2.5\sigma$.  At both densities, increasing the number of test particle insertions does little to help the inversion since it does not resolve the problem of lack of information.
At high temperature, $k_{\rm{B}}T/\varepsilon = 10$ [Fig.~\ref{fig:shoulder}(b) and (d)], TPI (blue line) is much more successful at both densities since, similar to Fig.~\ref{fig:comp-simu}(e) and (f), the temperature is high enough that the information issue is removed: more particles are able to overcome the barrier into the soft shell.  Despite the thermal energy being so much higher than the shoulder, the effect of the shoulder is evidently not washed out in $g^{*}(r)$ and can still be recovered.

As with the TPI method, IBI (olive-green line) works correctly at high temperature [panels (b) and (d)] and allows us to extract the potential, albeit with some fluctuations.  However, for low temperature [panels (a) and (c)], the potential from IBI does not resemble $u_{\rm{}s}(r)$ in the low-density case.  IBI struggles where $g^*(r) \approx 0$ as it has no information on which to base the inversion.  In practice, we can only assume that the potential is much higher than the thermal energy, i.e., $u(r_i)\gg k_{\rm B}T$, at points where $g^{*}(r_i)=0$.  In sparsely sampled regions, this results in the IBI potential making large jumps when completely empty bins are encountered in the target RDF.  While such gaps can, in principle, be reduced by a more thorough sampling of $g^{*}(r)$, it is not always possible to do so in practice.
The higher density case [panels (c) and (d)] suffers less from this effect as more information is available between the first and second peaks in $g^*(r)$. Nevertheless, with a shoulder height that is significantly higher than the energy available to the particles, there is poor sampling at the rise of the shoulder itself as particles may move freely within the flat part of the soft core but less easily in the steep section that leads onto it.  
Overall, IBI (olive-green lines) recovers the shoulder region more accurately than TPI  (blue lines) under these conditions although the shoulder incorrectly has a slight downward slope towards the hard core, producing a shallow well there [panel (c)].

In contrast to IBI and TPI, the ML method (red line) provides quite reliable information about the potential at all four combinations of temperature and density.  There are small discrepancies in the low-temperature case for both densities, most noticeably at low density [Fig.~\ref{fig:shoulder}(a)], where the repulsive core is too gentle.  Overall, ML finds similar information to that found by TPI (blue line) for all cases except the low-temperature, high-density case, where ML performs decisively better than both TPI and IBI.

\subsection{Experimental colloids}

\begin{figure}[h!]
\centering
\includegraphics[width=\linewidth]{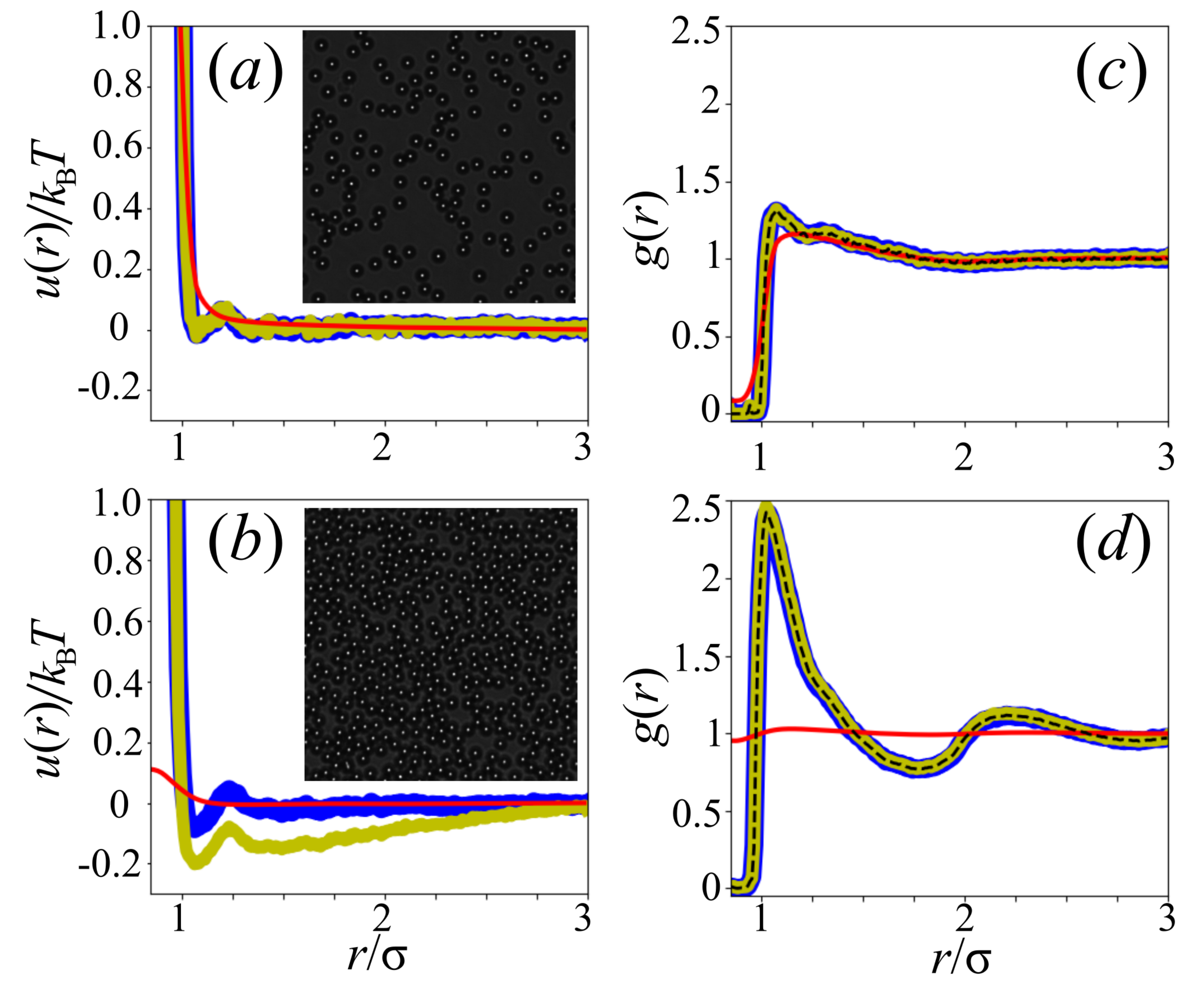}

   \caption{Experimental results for the three inversion methods: (i) TPI (blue), (ii) IBI (olive-green) and (iii) ML (red). We report  interaction potentials with inset example sections of images of (a) low and (b) high density MF particles of diameter 2.84 $\mu \mathrm{m}$. Also shown [panels (c) and (d)] is a comparison of $g^*(r)$ (black dashed line) computed from the particle coordinates with $g(r)$ obtained by TPI from each inverted $u(r)$ [color code as in panels (a) and (b)]. }
   \label{fig:experiments}
\end{figure}

Figures \ref{fig:experiments}(a) and (b) present the pair potentials obtained from each method by inverting the experimental data at low and high densities, respectively. Unlike the synthetic data from simulations of WCA, LJ, and the shoulder potential, we do not know the underlying potential {\it a priori}. However, we can compare the results between the three inversion methods and comment on the physical interpretation of the resulting potentials. The corresponding RDFs in panels (c) and (d) will be compared and discussed in Section \ref{sec:comparison}.

As shown in the figure, we obtain effective pair potentials at each density.  TPI [blue curves in panels (a) and (b)] returns pair potentials approximating hard spheres in each case (low and high density), with some small oscillations in $u(r)$ around $r = 1.2 \sigma$.  The exact shape of these oscillations differs slightly between the two sets of results, but the  scale of the oscillations is less than $0.1 k_{\rm{B}}T$.  It is expected that the potentials are similar at the different densities since the same particle type is being studied.  However, it is not surprising that some slight differences arise between the two densities, for a number of reasons.  First, it is difficult to exactly replicate the same experimental conditions; for example, the different particle densities might have slightly different background salt concentrations.  Second,  particles are not perfectly monodisperse (the coefficient of variation is 2.4\%); an example of this can be seen in the bottom left corner of the inset image in Fig.~\ref{fig:experiments}(b).  Third, particles may occasionally stick together or to the glass bottom cell.  Aside from these experimental reasons, there may also be higher-order interactions that only become apparent at increasing density.\cite{stones2023}  We have assumed that only pairwise interactions are present, so any higher-order interactions will appear as density-dependent differences in the effective pair potential.

The inverted potential from IBI [olive-green curves in panels (a) and (b)] is in excellent agreement with that from TPI (in blue) for the low-density case [panel (a)], providing confidence in the result. However, at high density [panel (b)], while having a similar shape, the potential found by IBI lies up to $0.2k_{\rm{B}}T$ below the potential obtained with TPI. The discrepancy gradually decreases as $r$ increases, with $u(r)$ from both IBI and TPI approximately reaching zero by $r = 3\sigma$ (beyond which we have set $u(r)=0$).  Hence, IBI infers a shallow attractive well from the high-density data.  We expect that the TPI results are more likely to be correct, given their similarity with the low-density result.
For the high-density case, IBI converged to a similar level as the low-density case in $\sim60$ iterations, which is consistent with performance on the results generated by simulation in Secs.~\ref{sec4a}--\ref{sec4c} at a similar density.

At low density, $u(r)$ from ML [red curves in panels (a) and (b)] follows the general shape of the potentials found by IBI and TPI, identifying the steep core potential of the hard sphere. However, the potential from ML lacks the finer resolution in $u(r)$ found by IBI and TPI.
At high density, $u(r)$ from ML follows the general shape of the potential found by TPI for $r > \sigma$, predicting approximately zero inter-particle interaction at these distances. However, $u(r)$ for $r < \sigma$ is qualitatively incorrect, failing to return a hard-sphere repulsion. As in the case of the high-density simulations, this may be due to the fact that the correlation between positions and velocity is lost more quickly at high density.  The problem becomes even more challenging when the repulsion approaches that of a perfect hard sphere, since the correlation between velocities and forces disappears.  Hence, the demands on precision and frequency of data acquisition rise steeply (see the supplementary material, Sec.~S3).

\section{Discussion}\label{sec:comparison}

\subsection{Mean absolute error as a measure of accuracy}\label{sec:MAE}

\begin{figure}[h!]
   \centering
   \includegraphics[width=0.8\linewidth]{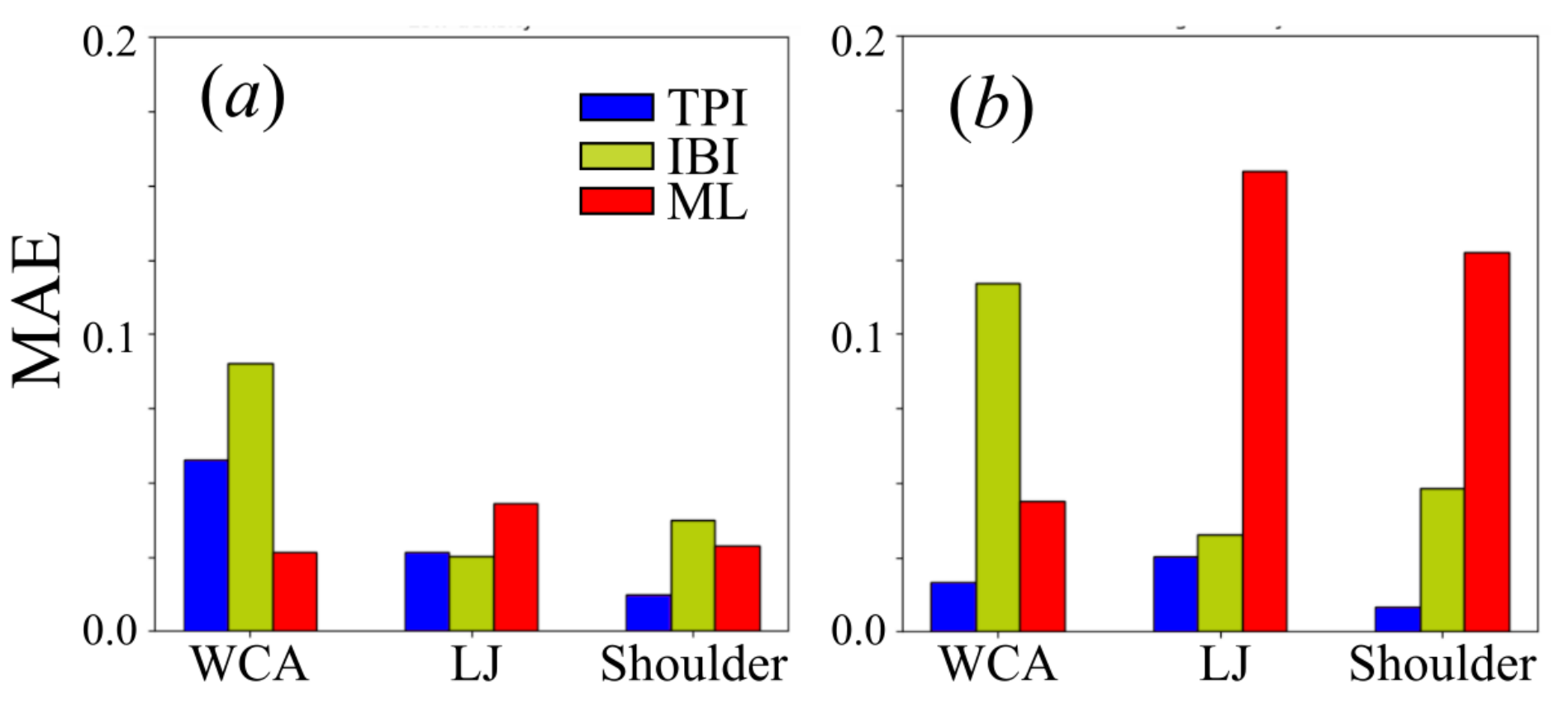}
   \caption{The mean absolute error for each method and potential presented in Fig.~\ref{fig:comp-simu} of the main text: (a) low density and (b) high density.}
   \label{fig:potentialMAE}
\end{figure}

As a quantitative means of comparing the results of the three methods for $u(r)$, we calculate the mean absolute error between the numerical result for the potential, $u^{\text{(n)}}(r)$, and the theoretical (i.e., exact) expression $u^{\text{(t)}}(r)$,
\begin{equation}
   \text{MAE}= \frac{1}{N_{\rm{p}}}\, \sum_i \left|u^{\text{(n)}}(r_i)-u^{\text{(t)}}(r_i)\right|,
\end{equation}
where $N_{\rm{p}}$ is the number of $r_i$ values evaluated in the range $0.9\, \sigma<r_i<3 \sigma$ and the sum is restricted to the same range.  In Fig.~\ref{fig:potentialMAE}, we report the computed MAE for all the examples from Fig.~\ref{fig:comp-simu}.

TPI obtains low MAE values in each case, reflecting the good agreement seen with the analytical potentials in Fig.~\ref{fig:comp-simu}. In most cases, TPI obtains the lowest MAE values of the three methods, except for LJ low density, for which TPI's MAE is just slightly higher than that of IBI.  The kinks in the IBI $u(r)$ at $r \sim 0.9 \sigma$ are one of the contributions that make IBI's MAE higher. Nevertheless, IBI also generally shows low MAE values.

For the ML method, the WCA potential is reproduced quite accurately although for the high density case, the MAE is somewhat higher. The reason is that ActiveNet bases its prediction on the correlations in the system, in particular between the instantaneous displacement (mean velocity) and the relative position between particles. In a dense system, the correlations between these quantities disappear sooner than in a dilute one, giving a worse result. The discrepancies in ML's $u(r)$ for the high-density LJ and shoulder potential cases, as seen in Fig.~\ref{fig:comp-simu} [panels (d) and (f)], are reflected in their high MAE.

\subsection{Test particle insertion as a measure of accuracy}\label{sec:grTPIaccuracy}

Perhaps a more stringent way to compare the inversion accuracy of IBI, TPI and ML for the simulated systems is by using the converged pair potentials to regenerate the RDFs.  There is a choice of method for finding the RDFs; starting from any $u(r)$, the corresponding $g(r)$ can be obtained from MC or BD simulations---both in conjunction with a distance histogram---or by TPI sampling using Eq.~(\ref{eqnTPI}).  We will first compare the RDFs from all three inversion methods using a single RDF method.  We choose TPI sampling for this purpose, noting that Eq.~(\ref{eqnTPI}) is exact and avoids the need to run new simulations to obtain the RDFs.

The TPI sampling is performed using an evenly spaced grid of insertion points with the same resolution as described in Section \ref{subsecTPI}, i.e., four times the number of sampling points as particles in the system.  For fairness, we have shifted the grid of insertion points by half the grid-spacing in each dimension when testing the potential from TPI inversion itself so that the predictions of the inversion process will not be tested using the same samples that were used to create it in the first place.

\begin{figure}[h!]
   \centering
   \includegraphics[width=0.9\linewidth]{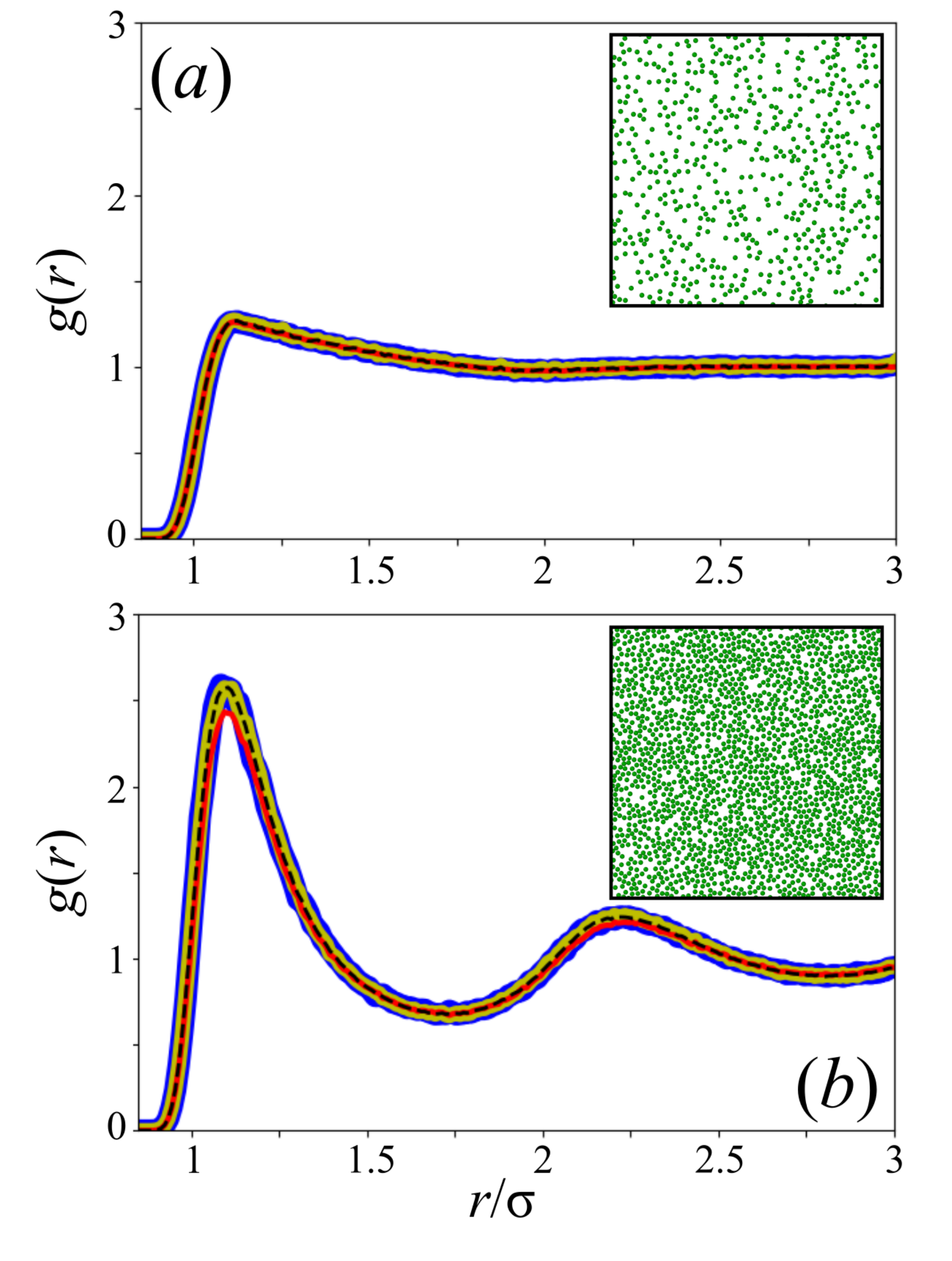}
   \caption{Radial distribution function found using TPI for the different numerical potentials computed with each method: TPI results are in blue, IBI in olive-green, and ML in red. Panel (a) reports results for WCA, $\rho =0.2$, and panel (b) reports results for LJ, $\rho=0.6$. The black dashed line represents the target $g^*(r)$ computed from the original BD data used for all the methods. Inset: example snapshots of particle configurations. 
   \label{fig:gdr}
   }
\end{figure}

The comparisons are plotted in Fig.~\ref{fig:gdr} for the low-density WCA [panel (a)] and high-density LJ [panel (b)] cases, using the potentials $u(r)$ presented in Fig.~\ref{fig:comp-simu}(a) and (d).  In the low-density WCA case [Fig.~\ref{fig:gdr}(a)], all the inverse methods return a pair potential that faithfully reproduces the target RDF as originally generated by the exact potential $u_{\rm WCA}(r)$ (black dashed line).  However, for the high-density LJ case [Fig.~\ref{fig:gdr}(b)], $u(r)$ from IBI and TPI produce RDFs that agree well with the target (black dashed line), while ML shows some discrepancy around the peak at $r = 1.1\sigma$.

\begin{table*}
\caption{\label{tab:comparison}Summary of the comparison of the three inversion methods: iterative Boltzmann inversion (IBI), test-particle insertion (TPI), and ActiveNet machine learning (ML).}
\begin{ruledtabular}
\begin{tabular}{cp{3cm}p{3cm}p{4.5cm}p{4.5cm}}
Method&Requirements/Input&Natural output&Advantages&Disadvantages\\ \hline
IBI&
$g(r)$&$u(r)$&
   \begin{itemize}[leftmargin=*,topsep=0pt]
   \vspace{-0.2cm}
       \item Only RDF or structure factor needed, no particle coordinates.
       \item Low memory/storage requirements.
   \end{itemize}&
   \begin{itemize}[leftmargin=*,topsep=0pt]
   \vspace{-0.2cm}
       \item Can struggle to equilibrate at high densities to obtain trial $g(r)$.
       \item Well-converged MC simulations take relatively long time.
       \item Struggles when $g^*(r) \rightarrow 0$.
       \item Sensitive to noise in target $g^*(r)$.
       \item Derived for equilibrium states.
   \end{itemize}\\
TPI&
Snapshot(s) of particle coordinates (these do not need to be successive). Independent frames are preferred. &$u(r)$&
   \begin{itemize}[leftmargin=*,topsep=0pt]
   \vspace{-0.3cm}
       \item Only uses the original particle coordinates, so does not require further simulations and forces convergence to the original phase.
       \item Can use more insertion points to make up for fewer snapshots.
   \end{itemize}&
   \begin{itemize}[leftmargin=*,topsep=0pt]
   \vspace{-0.3cm}
       \item At too high density, difficulty inserting successfully leads to poor/noisy inversion (but the densities studied here are fine).
       \item Struggles when $g^*(r) \rightarrow 0$.
       \item Sensitive to noise in target $g^*(r)$.
       \item Derived for equilibrium states.
   \end{itemize}\\
ML
&Time-ordered snapshots of particle coordinates, sufficiently close to track the particles.&$\rm{\textbf{F}}(\textit{r})$, which can be transformed into a potential by integration.&
   \begin{itemize}[leftmargin=*,topsep=0pt]
   \vspace{-0.3cm}
       \item No iterative scheme required.
       \item Easily generalizable for application to more complex potentials with other dependencies.
       \item Does not require an equilibrium state.
   \end{itemize}&
   \begin{itemize}[leftmargin=*,topsep=0pt]
   \vspace{-0.3cm}
       \item Requires velocities of particles, estimated from correlated snapshots.
       \item Requires a large amount of data to determine interactions.
       \item Sensitive to the correlation between velocities and interaction force.
   \end{itemize} \\

\end{tabular}
\end{ruledtabular}
\end{table*}

Arguably, inverting all potentials using TPI sampling is unfair to the ML approach, which returns forces $F(r)$ rather than $u(r)$, requiring integration before TPI can be applied.  The integration step is a source of error in $u(r)$ and, hence, also in $g(r)$ when calculated by TPI sampling.  In the supplementary material Sec.~S2, we compare the RDFs on a different basis, where each method is tested using its own approach: $g(r)$ for IBI is found by the distance-histogram method using the coordinates from the MC simulation in the final iteration;  $g(r)$ for TPI is found using TPI, with the {\em same} insertion points as used in the inversion; and $g(r)$ for ML is found by using $F(r)$ directly in a BD simulation, then using the distance-histogram method on the resulting coordinates.  Each method's $g(r)$ now agrees well with the target $g^*(r)$.  This confirms that the apparent error in the ML $u(r)$ in cases such as the high-density LJ plot [Fig.~\ref{fig:gdr}(b)], is likely due to the numerical integration of $F(r)$, rather the inversion for $F(r)$ itself. To obtain accurate $u(r)$ values from ML, a more sophisticated process would need to be developed. This could involve, for example, performing symbolic regression for $F(r)$ to facilitate accurate integration. 

We now perform the equivalent comparison of RDFs using the experimental system.  Figures~\ref{fig:experiments}(c) and (d) compare the experimental target $g^*(r)$ with the RDF generated using TPI sampling of the potentials obtained by each inverse method. For the low-density experiment [panel (c)], we see excellent agreement between IBI, TPI, and the target $g^*(r)$, corresponding with the good agreement between the inverted potentials $u(r)$ in panel (a). The RDF computed from the ML $u(r)$ approximately follows the shape of $g^*(r)$, albeit somewhat smoothed, as follows from the corresponding $u(r)$.  This test suggests that the potential from ML is less accurate than that from IBI and TPI in this case.

In the high-density experiment, we see excellent agreement between $g^*(r)$ and TPI.  The IBI potential also yields a RDF that closely resembles $g^*(r)$, with only a slight overprediction of the first peak in $g(r)$.  The similarity of the RDFs from the TPI and IBI potentials contrasts with the systematic deviations between the potentials themselves, as shown in Fig.~\ref{fig:experiments}(b).  This suggests that the disagreement between the potentials is due to the sensitivity of the inversion process from $g(r)$ to $u(r)$ in IBI\cite{wang2020} despite the formal uniqueness of the mapping.\cite{henderson1974}

\subsection{Computational efficiency}
It is not straightforward to compare the computational efficiency of the inversion methods quantitatively due to the different nature of the numerical procedures (such as MC for IBI compared to GNN training for ML), the choice of implementations (several codes are available for some parts of the algorithms), and the degrees of freedom in parameter space (such as the number of frames, number of particles per frame, and number of test particles).

It is also important to note that changing a given parameter can have a different impact on the computational cost of each method.  A good example is the inversion cutoff, $r_{\rm{a}}$, which is required for all three methods.  We always require the cutoff to be at least as large as the true interaction range to obtain an accurate result for $u(r)$, but the true range is not always known in advance.  In IBI, an increase in $r_{\rm a}$ means that more pairwise interactions must be evaluated in each step of the MC simulation that informs the potential update step in each iteration.  Any cell list being used to improve the efficiency of the simulations must be made correspondingly coarser, thereby eroding the benefit of using the list.  In ML, $r_{\rm a}$ controls the number of pairs of connected particles in the GNN, for each of which the correlations then need to be tracked and computed.  Therefore, the value of $r_{\rm a}$ also has implications for the amount of memory needed for the ML analysis, especially in high-density systems.  TPI is less computationally intensive, so larger cutoffs can be accommodated more easily.  Although more distances between particles and test particles would need to be calculated if $r_{\rm{a}}$ is increased, these distances only need to be calculated once if the approach of fixed insertion points is adopted.

\section{Conclusions\label{sec:conclusion}}

We conclude with a commentary on the advantages and disadvantages of the three inverse methods, as summarized in Table~\ref{tab:comparison}. The overarching factors determining when each inverse method should be used concern the type of data available, and the intended use of the inversion result.

The ML approach is the most demanding in terms of input data, as it requires time-correlated snapshots. This point is particularly relevant to experiments, where it is necessary first to determine the minimum frame rate of the images to provide sufficient correlation.  On the other hand, ML is the only method that does not rely on equilibrium statistical distributions for its derivation, meaning that it is readily applicable in active matter and other non-equilibrium applications.

At the opposite extreme in terms of detail needed from the input data, IBI is the only choice for inverting a structure factor from a scattering experiment, i.e., with no data for the coordinates of individual particles.  TPI is likely the best choice when particle coordinates are available, at least until the ML approach is further refined; in the experimental example reported in this work, TPI gives the most reliable inverted pair potential under different conditions.  Furthermore, TPI is the least computationally intensive method when used with fixed insertion points.   Using the coordinates of all particles, not just the overall RDF, TPI is less prone to the sensitivity of $u(r)$ to small changes (or errors) in $g(r)$ which tend to affect IBI adversely.  At high particle densities, all methods start to break down but for different reasons.

Due to numerical limitations in integrating the forces $F(r)$, the current setup of ML is less attractive than IBI or TPI if the inverted pair-interaction $u(r)$ is needed for the intended application, such as MC simulations. Conversely, due to the difficulty in numerically differentiating $u(r)$, ML may be preferable for dynamics-based applications, such as BD simulations. A major advantage of ActiveNet is the ability to decouple one-body and two-body (pairwise) contributions---a feature that has not been put to the test in the present article.

There are some obvious but challenging directions in which this work could lead.  We have compared simulations and experiments for two-dimensional systems because it is more difficult to obtain three-dimensional structures by microscopy, where the additional depth coordinate must be determined accurately. Nevertheless, all the methods in this work extend naturally to three dimensions, and we expect the same conclusions to be drawn regarding the analysis of simulations. It is also of interest to apply the methods to many-body interactions,\cite{stones2023} multicomponent systems,\cite{thorneywork2014} and anisotropic interactions.\cite{newton2017,lei2023,kern2003,reynolds2015}  Work is in progress on these extensions.

One could also consider protocols that couple some of the methods to combine their complementary strengths.  A plausible improvement to IBI could be to first use ML to generate a closer initial guess of the potential. Indeed, more recently, IBI has been used to correct existing machine learning potentials,\cite{Matin2024} allowing for experimental data (via the RDF) to be incorporated to improve the accuracy of molecular dynamics simulations.  Hence, potential inversion methodology is still advancing.

\section{Supplementary material}

The first section of the supplementary material (S1) reports the effect of  choosing fixed vs random test particle coordinates in TPI.
The second section of the supplementary material (S2) further compares the three methods, computing $g(r)$ using the converged potential $u(r)$ via Monte Carlo (IBI), using particle
coordinates and the same grid of insertion points as was used in the optimization (TPI) or  Brownian Dynamics (ML).
Finally, the third section (S3) discusses the optimal velocity-force correlation in the ML method.

\begin{acknowledgments}

C.V. acknowledges funding from MINECO  (Grant Nos.~IHRC22/00002  and PID2022-140407NB-C21). C.R.R.Z. is funded by a Junior Research Fellowship from Christ Church, University of Oxford. D.E. acknowledges funding from the EPSRC SOFI$^2$ Center for Doctoral Training (Grant No. EP/S023631/1). J.M. acknowledges financial support from the UCM predoctoral contract (call CT15/23).

We acknowledge C. Miguel Barriuso-Guti{\'e}rrez and Miguel Ruiz-Garcia (Universidad Complutense de Madrid) and Luca Ghiringhelli (FAU Erlangen-N{\"u}rnberg) for help provided at the beginning of this project on adapting ActiveNet as used here to obtain the results of the ML method. This research made use of the Hamilton high-performance computing (HPC) facility at Durham University.

\end{acknowledgments}

\section*{Data Availability Statement}

Raw data for the figures in the main text are provided in a zip file in the Supplementary Information.

\section*{References}

\nocite{*}
\bibliography{aipsamp_AFTER_REVISION}

\end{document}

\preprint{AIP/123-QED}

\title[Numerical methods for unraveling inter-particle potentials]{Supplementary material: Numerical methods for unraveling inter-particle potentials in colloidal suspensions: A comparative study for two-dimensional suspensions}
 
\title{Supplementary information:\\ Numerical methods for unraveling inter-particle potentials in colloidal suspensions: A comparative study for two-dimensional suspensions}

\author{Clare R.~Rees-Zimmerman}
\affiliation{Physical and Theoretical Chemistry Laboratory, University of Oxford, South Parks Road, Oxford OX1 3QZ, United Kingdom}
\author{Jos{\'e} Mart{\'\i}n-Roca}
\affiliation{Departamento de Estructura de la Materia, F{\'\i}sica T{\'e}rmica y Electr{\'o}nica, Universidad Complutense de Madrid, 28040 Madrid, Spain}
\altaffiliation[Also at ] {Grupo Interdisciplinar Sistemas Complejos, Madrid, Spain}
\author{David Evans}
\affiliation{Department of Chemistry, Durham University, South Road, Durham DH1 3LE, United Kingdom}
\author{Mark A.~Miller}
\affiliation{Department of Chemistry, Durham University, South Road, Durham DH1 3LE, United Kingdom}
\author{Dirk G.~A.~L.~Aarts}
\affiliation{Physical and Theoretical Chemistry Laboratory, University of Oxford, South Parks Road, Oxford OX1 3QZ, United Kingdom}
\author{Chantal Valeriani}
\affiliation{Departamento de Estructura de la Materia, F{\'\i}sica T{\'e}rmica y Electr{\'o}nica, Universidad Complutense de Madrid, 28040 Madrid, Spain}
\altaffiliation[Also at]{Grupo Interdisciplinar Sistemas Complejos, Madrid, Spain}
\email{cvaleriani@ucm.es}

\begin{titlepage}
\maketitle
\end{titlepage}

\onecolumngrid

\renewcommand\thesection{S\arabic{section}}

\section{TPI: Fixed or random test particle coordinates}\label{appendix_tpi}

\begin{figure} 
    \centering
\includegraphics[width=0.85\textwidth]{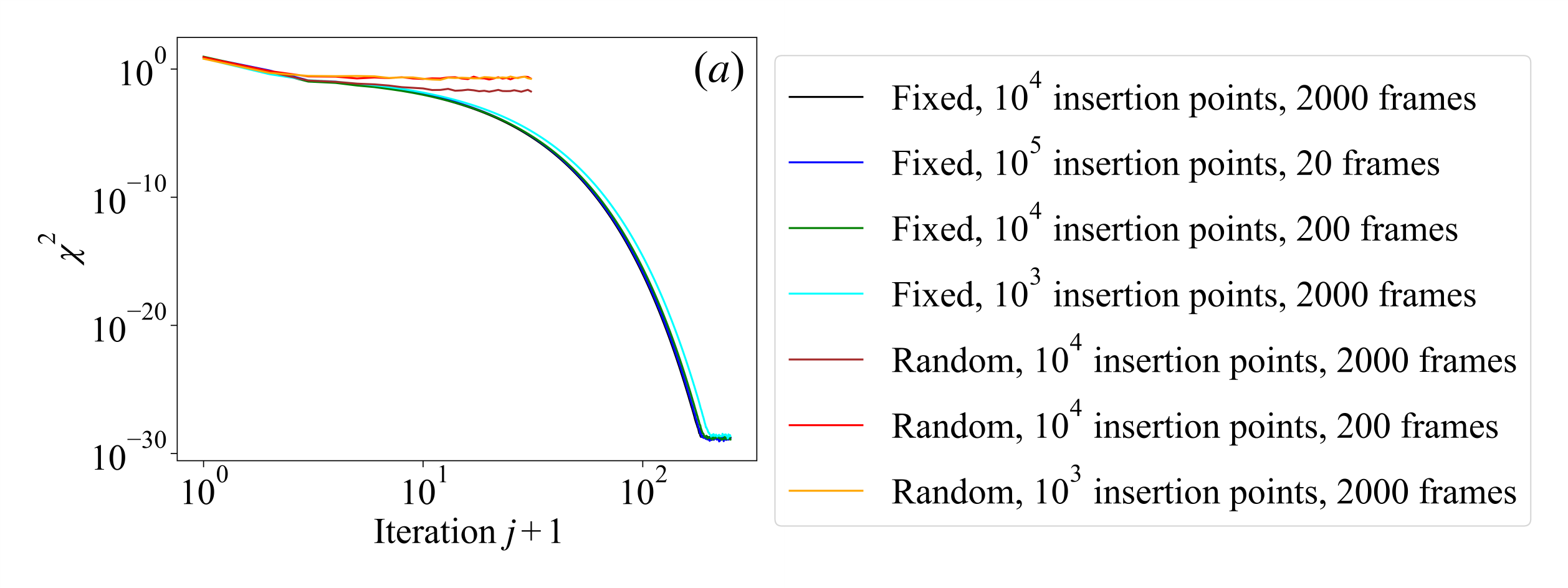} 
    \quad
    \break \hfill
\includegraphics[width=0.436\textwidth]{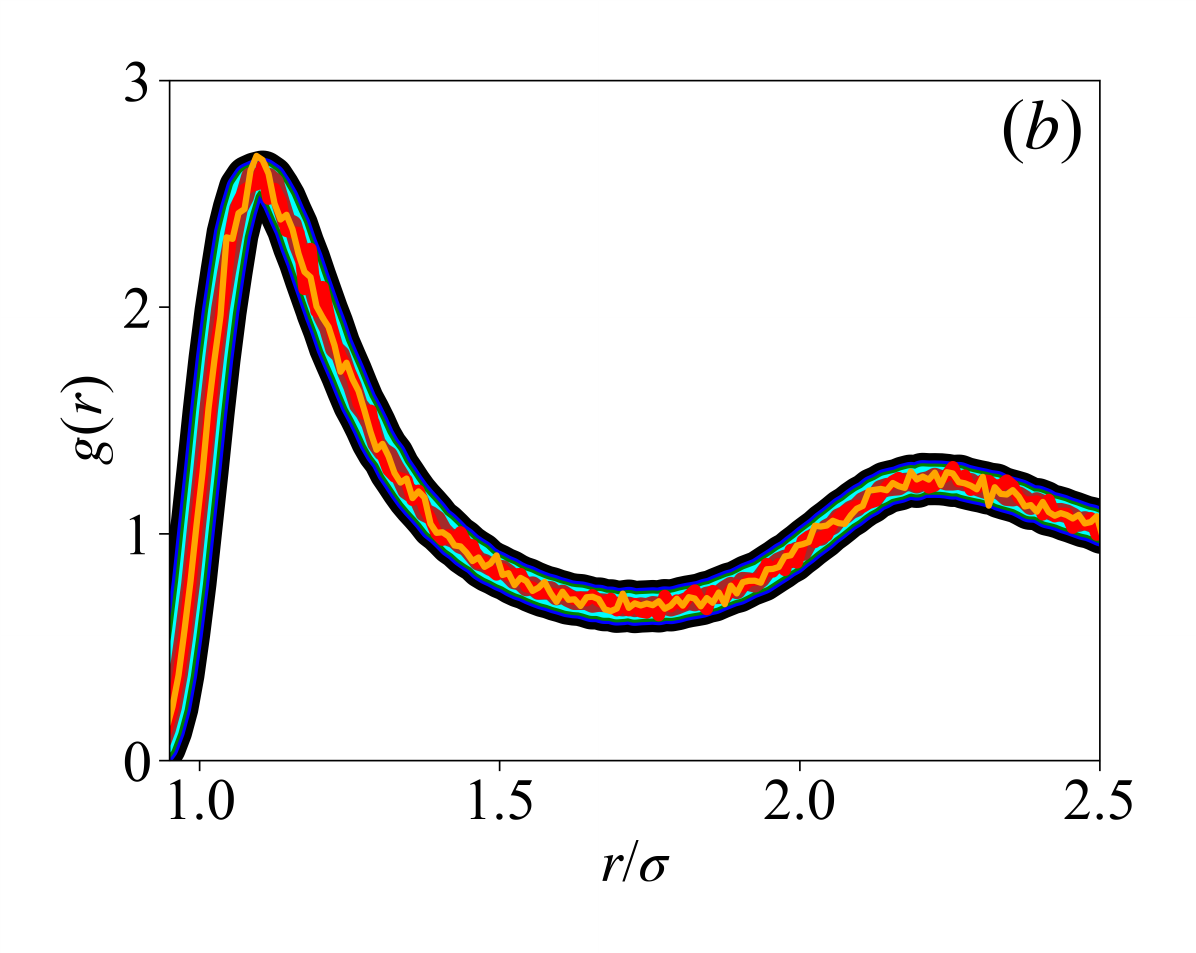}
\includegraphics[width=0.45\textwidth]{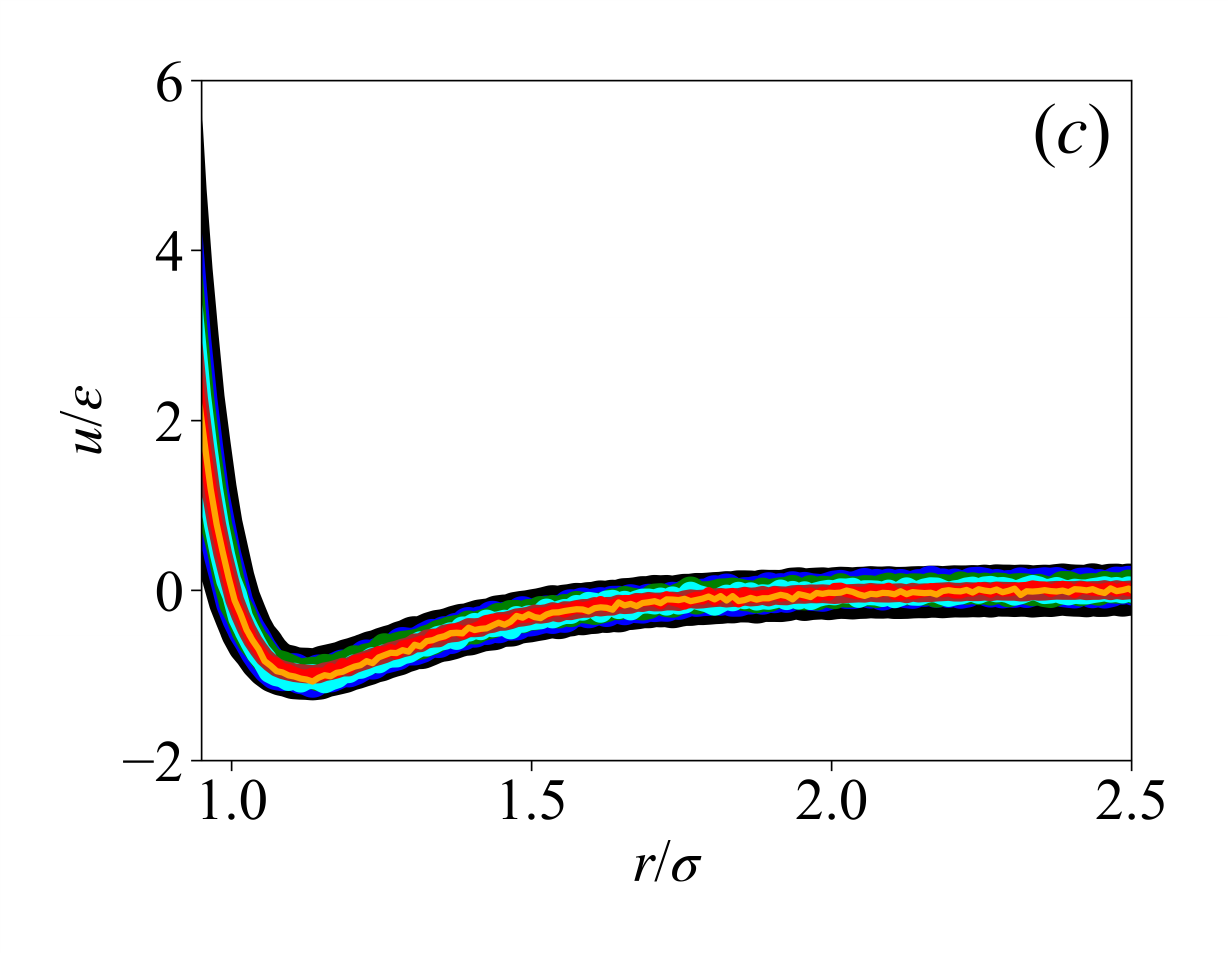}
    \caption[]{Convergence of TPI for LJ particles at reduced density $\rho=0.6$ for different combinations of insertion points per frame and number of frames (simulation snapshots).  (a) Mean squared deviation $\chi^2$ of the RDF from the target as a function of the TPI iteration number; (b) $g_{\rm{TPI}}(r_i)$ against $r/\sigma$; and (c) $u(r_i)$ against $r/\sigma$, for fixed (black/blue/green/cyan) and random (brown/red/orange) insertion points, with various combinations of number of insertion points and number of frames.}
    \label{fig:tpi_figure}
\end{figure}

To compare the impact of choosing fixed or random test particle coordinates, we monitor convergence via $\chi^2 = \sum_i \left[g_{\rm{TPI}}(r_i) - g^{*}(r_{i}) \right]^2$. For the example case of the LJ simulation at $\rho=0.6$, Fig.~\ref{fig:tpi_figure}(a) plots $\chi^2$ against $\left(j+1\right)$, where $j$ is the iteration number. For fixed insertion points, i.e., the same insertion points used in each iteration, it can be seen that $\chi^2$ keeps decreasing until it levels off at $\chi^2\sim10^{-29}$ after $\sim200$ iterations. We show one version with $2\times10^7$ total insertion points ($10^4$ insertion points$\,\times \,\, 2000$ frames), and three with $2\times10^6$ total insertion points (different combinations for number of insertion points and frames). These all give similar results, as plenty of insertion points are included; it makes little difference whether more frames or more insertion points per frame are used to attain the same total number of insertion points. However, of the versions with $2\times10^6$ total insertion points, convergence is slightly faster for that with $10^5$ points per frame than that with $10^4$, which in turn converges slightly faster than the $10^3$ version. The version with the greatest number of total insertion points $2\times10^7$ converges the fastest of them all.

Fig~\ref{fig:tpi_figure}(b) shows the corresponding converged $g_{\rm{TPI}}(r_i)$; for the fixed insertion points, this matches $g^{*}(r_{i})$ to machine precision. Calculated using these fixed test particle locations, noise levels are similarly small in $g_{\rm{TPI}}(r_i)$ for all of the combinations of number of insertion points and frames -- this is because they are matching the noise in $g^{*}(r_{i})$. Fig.~\ref{fig:tpi_figure}(c) plots $u(r_i)$. For the fixed insertion points, there is little noise in $u(r_i)$; it corresponds with the noise in $g_{\rm{TPI}}(r_i)$.

We repeat this study with new random insertion points generated for each iteration. In Fig.~\ref{fig:tpi_figure}(a), we observe that $\chi^2$ initially follows a similar decrease as with the fixed insertion points, but it then levels off around a much larger $\chi^2$ value ($\chi^2\sim10^{-1}$ with $2\times10^6$ total insertions and $\chi^2\sim10^{-2}$ with $2\times10^7$ total insertions). As a result, with random points, significantly fewer iterations are required for convergence. This implies that the small $\chi^2$ values obtained with the fixed insertion points were specific to the locations of those insertion points. Again, there is little difference between different combinations that form the same number of total insertion points. We see that an impracticably large number of total insertions would be required to decrease $\chi^2$ much further with random insertion points.

In Fig.~\ref{fig:tpi_figure}(b), there is more noise in $g_{\rm{TPI}}(r_i)$ with random insertion points than with fixed insertion points. This is because there is noise from imperfect sampling with test particles, in addition to any noise in $g^{*}(r_{i})$. Similarly, there is more noise in $u(r_i)$ with random insertion points than with fixed insertion points. It can be seen in Fig.~\ref{fig:tpi_figure}(c) that the noise in $u(r_i)$ decreases as the total number of insertion points increase from $2\times10^6$ to $2\times10^7$, due to improved sampling.

The main text of this work presents results with fixed insertion points due to the substantially increased computational efficiency: the biggest computational cost is calculating distances between particles and test particles, which would have to be repeated every iteration with random insertion points. Whilst the fixed insertion point codes are run until convergence to machine precision, it raises the question, at what point should we stop an optimisation using fixed points, if we wish to have a meaningful level of noise in the resulting $g_{\rm{TPI}}(r_i)$---i.e., one which does not depend on the choice of test particle coordinates?

It is worth mentioning that if the converged $u(r_i)$ is used to calculate $g_{\rm{TPI}}(r_i)$ for $r_i$ greater than the cutoff $r_{\rm a}$ that was used in the inversion, then, with random insertion points, the resulting noise in $g_{\rm{TPI}}(r_i)$ is the same both before and after the cutoff in $r_i$. With fixed insertion points, whilst we can reduce the noise in $g_{\rm{TPI}}(r_i)$ to match that in $g^{*}(r_{i})$ up to the cutoff, noise will be retained \textit{beyond} the cutoff. Approximately matching the noise in $g_{\rm{TPI}}(r_i)$ before and after the cutoff could provide a practical answer to the question of when to stop iterating.

\section{Further comparison of methods}\label{appendix_gr}

\begin{figure}[h!]
    \centering
    \includegraphics[width=0.95\linewidth]{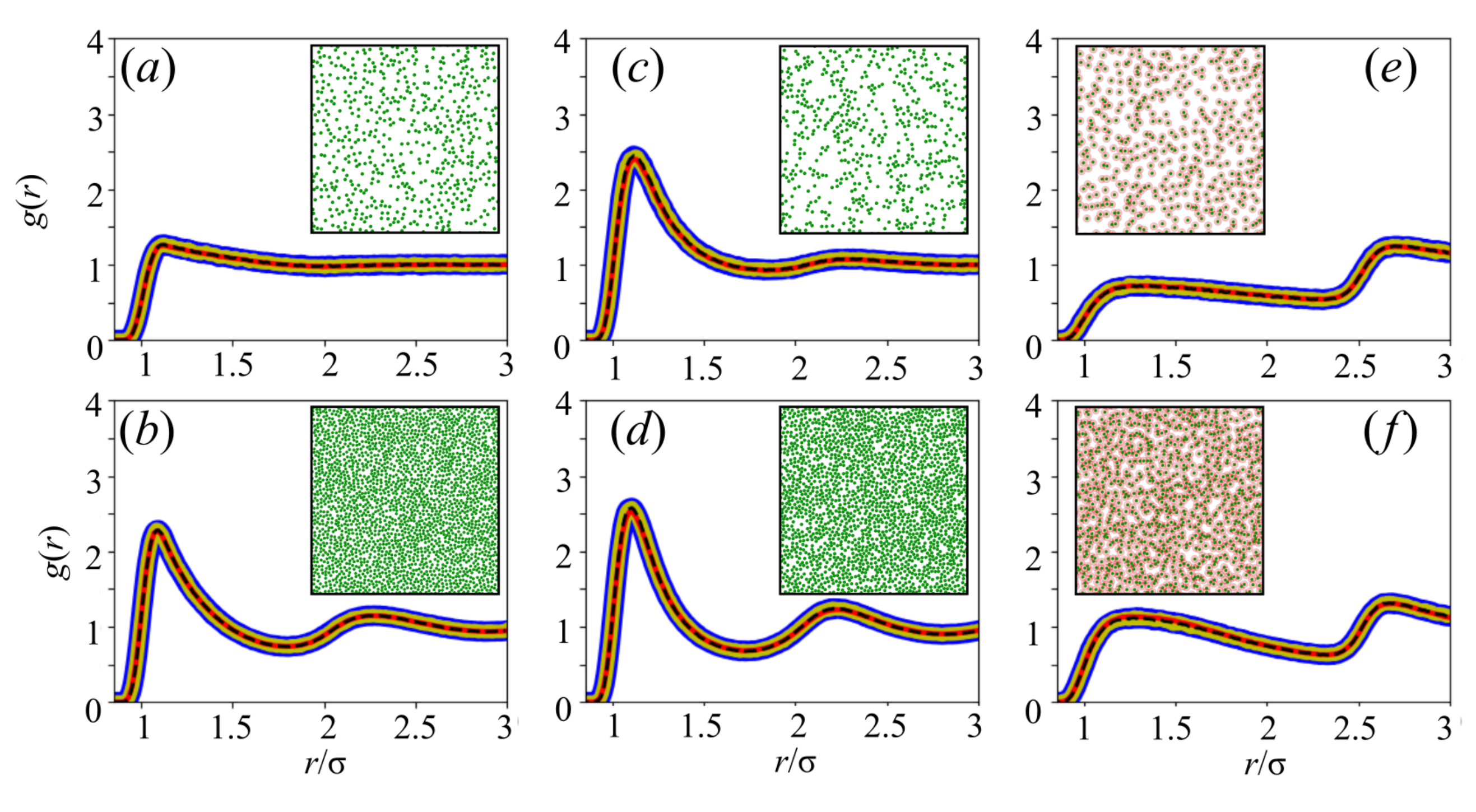}
    \caption{Radial distribution function derived by each method, for each of the numerical potentials computed. The panels represent the different potentials: WCA, (a) $\rho =0.2$ and (b) $\rho=0.6$;  LJ, (c) $\rho=0.2$ and (d) $\rho=0.6$; shoulder (e) $\rho=0.1$ and (f) $\rho=0.227$. The color of each curve represents the method used: IBI (olive-green), TPI (blue) or ML (red). The black dashed line represents the target $g^*(r)$ computed from the original BD data used for all the methods. Inset: example snapshots of particle configurations.}
    \label{fig:gdrmethods_SI}
\end{figure}

In Fig.~6 of the main text, the potentials $u(r)$ from each of the three inversion methods are compared by using them to calculate $g(r)$ using TPI sampling on the particle coordinates from the original simulations. As an additional means of comparison, here we calculate $g(r)$ for each inverse method, as evaluated by its own method: for IBI, we calculate $g(r)$ using the distance-histogram method in the MC simulation of the final iteration; for TPI, we calculate $g(r)$ using the converged $u(r)$, the particle coordinates, and the same grid of insertion points as was used in the optimisation; for ML, we run a BD simulation with the $F(r)$ that this method delivers.  In Fig.~\ref{fig:gdrmethods_SI}, we present all the simulation cases. Now, in all cases, we see excellent agreement between $g^*(r)$ and the evaluated $g(r)$ after the inversion. 

In particular, whilst Fig.~6 showed poorer agreement for ML in the high-density LJ case (Fig.~6(b)), if we instead test how well $F(r)$ regenerates $g^*(r)$, we get excellent agreement. This suggests that the disagreement in Fig.~6 was purely due to error from numerically integrating $F(r)$.  IBI and TPI still show similarly good agreement as is seen in Fig.~6, primarily as the testing methods are energy-based in each case. As discussed in SI Section \ref{appendix_tpi}, in Fig.~\ref{fig:gdrmethods_SI}, $g_{\rm{TPI}}(r)$ is overfitted due to reusing the same insertion points as were used in the optimisation. In Fig.~6 of the main text, whilst not being overfitted, due to using a different set of insertion points for the test, the agreement is still good enough that the difference is not visible by eye.

\section{ML: Optimal correlation time for learning}\label{appendix_ML}

\begin{figure}[h!]
    \centering
    \includegraphics[width=0.5\linewidth]{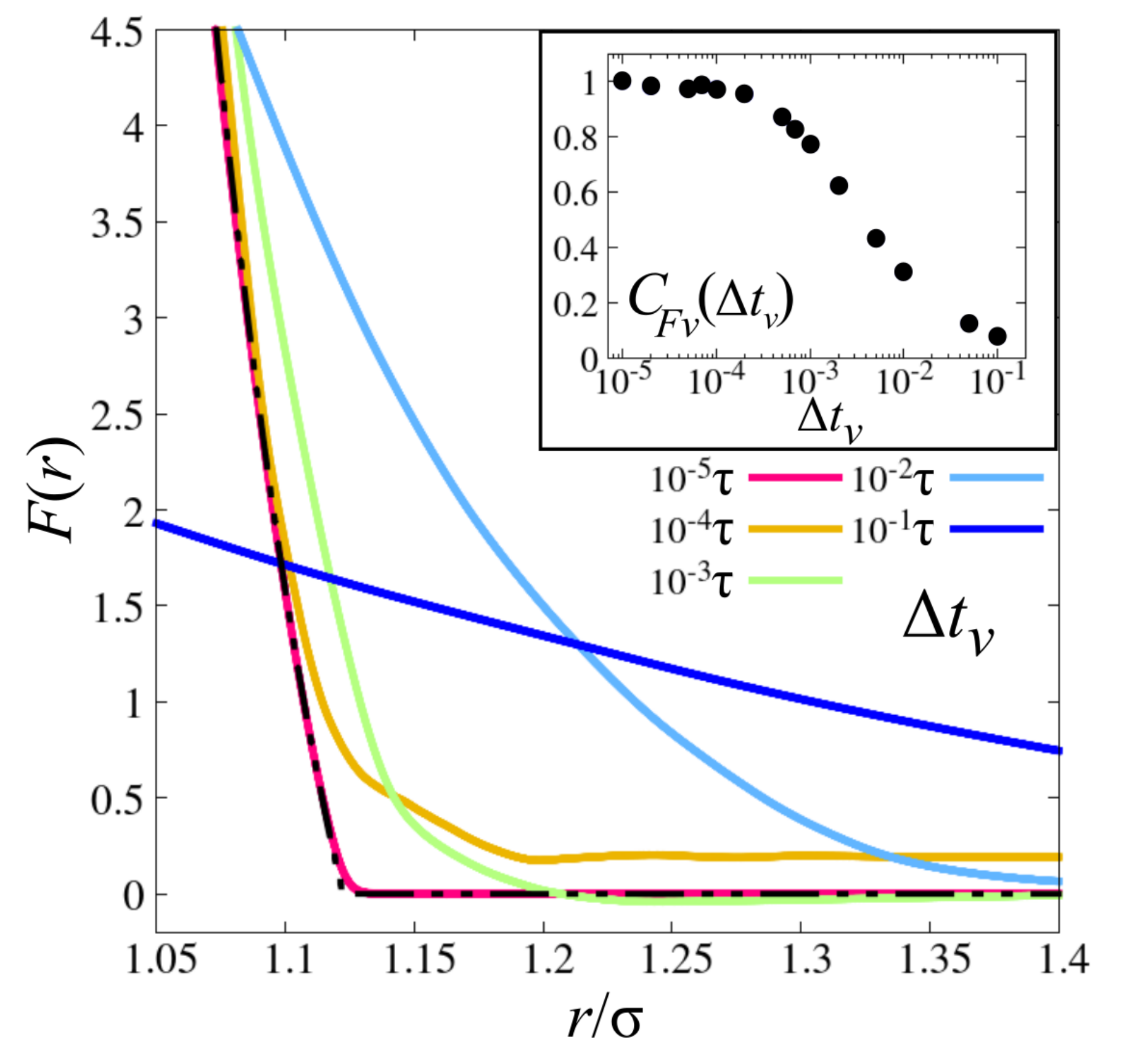}
    \caption{Predicted pair interaction force in a WCA simulation as a function of relative distance of particles, for different values of the time interval, $\Delta t_v$. The black dashed line presents the exact WCA force. Inset shows the force--velocity correlation as a function of $\Delta t_v$.}
    \label{fig:Corr}
\end{figure}

Since in Brownian Dynamics the instantaneous velocity is not properly defined (i.e., it is a random variable), the velocities used to train the network are deduced from the displacement between frames separated in time by $\Delta t_v$:
\begin{equation}
   \mathrm{\textbf{v}}^{\, \mathrm{measure}} (t,\Delta t_v) \approx \frac{\mathrm{\textbf{r}}(t+\Delta t_v) -\mathrm{\textbf{r}}(t)}{\Delta t_v}.
\end{equation}
For the ML method to learn the forces, these average velocities over the interval $\Delta t_v$ must be sufficiently correlated with the forces experienced by the particles.  This is because the network is looking for the relationship $\gamma \mathrm{\textbf{v}}=\mathrm{\textbf{F}}$. In Fig.~\ref{fig:Corr}, we show the prediction of the WCA potential at reduced density  $\rho=0.2$ using different values of $\Delta t_v$ to estimate the velocities from the displacements.  As $\Delta t_v$ increases, the predicted force deviates increasingly from the true WCA force.  In the inset of Fig.~\ref{fig:Corr}, we present the normalized time correlation function, $C_{Fv}(t)$, for the forces at one step $t_0$, with the velocities computed using different values of $\Delta t_v$,
\begin{equation}
    C_{Fv}(t) = \frac{\langle \mathrm{\textbf{F}}(t_0) \cdot \mathrm{\textbf{v}}(t_0+t) \rangle}{\langle \mathrm{\textbf{F}}(t_0) \cdot \mathrm{\textbf{v}}(t_0) \rangle},
\end{equation}
where $\mathrm{\textbf{F}}(t_0)$ is the forces acting on one particle at $t_0$ and $\langle ... \rangle$ represents the average over the ensemble and different initial configurations. For the values of $\Delta t_v$ when the two quantities are correlated, i.e., when $C_{Fv}(t)$ is close to 1, the prediction of the forces is much better than for high values of $\Delta t_v$.

The correlations between forces and velocities are easy to calculate in simulations.  However, it is not always possible to obtain such detailed information in experiments.  The measurement of this correlation provides a criterion for the parameters chosen in the main text to obtain a good prediction for the force.  Empirically, we find that the shape of the velocity autocorrelation function $C_{vv}$ (data not shown) is similar to that of $C_{Fv}$.  The similarity can provide a basis for estimating the accuracy of the ML inversion method might be before embarking on the analysis.